\documentclass[preprint, 3p]{elsarticle}
\journal{CAS}
\usepackage{graphicx}
\usepackage{amsmath}
\usepackage{amssymb}
\usepackage{xspace}
\usepackage{color}
\usepackage{subcaption}
\usepackage{longtable,booktabs}
\usepackage{lineno}
\newcommand{\nc}{\newcommand}
\nc{\rnc}{\renewcommand}
\nc{\bs}{\boldsymbol}
\rnc{\matrix}[2]{\left[\!\!\begin{array}{#1}
	#2\end{array}\!\!\right]}
\rnc{\vector}[1]{\matrix{c}{#1}}
\nc{\mm}[1]{\boldsymbol{#1}}
\nc{\mms}[1]{\boldsymbol{#1}}
\nc{\real}[1]{\Re\left\{ #1 \right\}}
\nc{\imag}[1]{\Im\left\{ #1 \right\}}
\nc{\dd}{\mathrm{d}}
\nc{\ii}{\mathrm{i}}
\nc{\ee}{\mathrm{e}}
\nc{\inv}{^{-1}} 
\nc{\herm}{^{\mathrm H}}
\nc{\tra}{^{\mathrm T}}
\nc{\conj}[1]{ \overline{#1} }
\nc{\diag}{\operatorname{diag}}
\nc{\CoR}{e}
\nc{\COMMENT}[1]{\textcolor{red}{#1}}
\nc{\ie}{i.\,e.\xspace}
\nc{\eg}{e.\,g.\xspace}
\nc{\cf}{cf.\,}
\nc{\myquote}[1]{`#1'}
\nc{\etal}{et al.\xspace}
\nc{\etc}{etc.\xspace}
\nc{\fabstand}{\,}
\nc{\fp}{\fabstand.}
\nc{\fk}{\fabstand,}
\nc{\tab}[5][tbh]{\begin{table}[#1]\centering\caption{#4\label{tab:#5}}\begin{tabular}{#2}\hline #3 \\ \hline\end{tabular}\end{table}}
\nc{\fig}[4][tbh]{
\begin{figure}[#1]
\centering
\includegraphics[width=#4\textwidth]{figures/#2}
\caption{#3\label{fig:#2}}
\end{figure}}
\nc{\e}[1]{\begin{equation} #1 \end{equation}}
\nc{\est}[1]{\begin{equation*} #1 \end{equation*}}
\nc{\ea}[1]{
\begin{eqnarray}
#1\end{eqnarray}}
\nc{\east}[1]{
\begin{eqnarray*}
#1 \end{eqnarray*}}
\nc{\fref}[1]{{Fig.~\ref{fig:#1}}}
\nc{\frefo}[1]{{\ref{fig:#1}}}
\nc{\frefs}[1]{{Figs.~\ref{fig:#1}}}
\nc{\tref}[1]{{Tab.~\ref{tab:#1}}}
\nc{\trefo}[1]{{\ref{tab:#1}}}
\nc{\trefs}[1]{{Tab.~\ref{tab:#1}}}
\nc{\eref}[1]{{Eq.~(\ref{eq:#1})}}
\nc{\erefo}[1]{(\ref{eq:#1})}
\nc{\erefs}[1]{{Eqs.~(\ref{eq:#1})}}
\nc{\sref}[1]{{Section~\ref{sec:#1}}}
\nc{\srefo}[1]{\ref{sec:#1}}
\nc{\srefs}[1]{{Sections~\ref{sec:#1}}}
\nc{\aref}[1]{{{Appendix~\ref{asec:#1}}}}
\nc{\arefo}[1]{{\ref{asec:#1}}}
\nc{\arefs}[1]{{{Appendices~\ref{asec:#1}}}}

\makeindex             

\begin{document}

\begin{frontmatter}
\title{A massless boundary component mode synthesis method for elastodynamic contact problems}
\author{Monjaraz Tec, C.D.$^1$}
\author{Gross, J.$^1$}
\author{Krack, M.$^1$}
\address{$^1$ University of Stuttgart, GERMANY}

\begin{abstract}
We propose to combine the ideas of mass redistribution and component mode synthesis.
More specifically, we employ the MacNeal method, which readily leads to a singular mass matrix, and an accordingly modified version of the Craig-Bampton method.
Besides obtaining a massless boundary, we achieve a drastic reduction of the mathematical model order in this way compared to the parent finite element model.
We assess the method's computational performance by a series of benchmarks, including both frictionless and frictional contact.
The results indicate that the proposed method achieves excellent energy conservation properties and superior convergence behavior.
It reduces the spurious oscillations and decreases the computational effort by about 1-2 orders of magnitude compared to the current state of the art (mass-carrying component mode synthesis method).
We believe that the computational performance and favorable energy conservation properties will be valuable for the prediction of vibro-impact processes and physical damping.
\end{abstract}


\end{frontmatter}
\section{Introduction}
\label{sec:introduction}
A vibro-impact process denotes any physical process where the mutual interaction between temporary (opening-closing) contacts and vibrations plays a central role \cite{Babitsky.2013}.
Technical examples are
machining processes (\eg pneumatic hammers, ultrasonic drillers),
impact vibration absorbers,
systems with backlash/freeplay (\eg hammering gears in drive trains, piping systems),
rotor-stator interactions (\eg between blades and casing of aero engines or within rotor bearings).
Vibro-impact processes have multi-scale character both in time and in space:
The contact time of a single impact is very short compared with the periods of the critical low-frequency vibration modes.
The contact region, in which high stresses are generated, is very small compared to the wavelengths of the critical vibration modes.
An elastic wave is initiated from the collision point, which then propagates within the solid body.
The wave decays due to dissipative effects within the solid and the boundaries.
Depending on the boundary conditions, the wave is reflected, travels back, and may thus affect the contact interactions \cite{Seifried.2010}.
In the presence of vibrations, recurrent impacts take place.
\\
To accurately predict vibro-impact processes, an appropriate combination of models of the elastodynamics within the bodies and the contact mechanics is required \cite{wrig2006,john1989,Ibrahim.2009}.
This poses a number of computational challenges, as explained in the following.

\subsubsection*{Enormous model order}
A fine space discretization is needed to describe the mechanics in the contact region accurately and to determine the stresses within the whole body.
The short collisions transfer considerable energy into high-frequency vibration modes.
The finite element model must properly resolve the shortest wavelength which is associated with the highest relevant modal frequency.
This yields an enormous mathematical model order, not seldom in the range of $10^5$–$10^7$ nodal displacement degrees of freedom in the case of three-dimensional problems.
As in the linear case, Component Mode Synthesis methods are popular in the vibrations community for model order reduction in the presence of nonlinear contact conditions.
Here, the idea is to approximate the displacement field in terms of a linear combination of component modes, and to project the momentum balance onto the subspace spanned by these modes (Galerkin procedure) \cite{klerk2008}.
As component modes, a subset of normal vibration modes can simply be used.
This corresponds to the classical modal truncation, which has been applied to impact problems in combination with regular contact models like Hertzian springs \cite{Seifried.2010}.
It is common to combine static deflection shapes obtained for imposed displacements or imposed forces at the contact interface with fixed- or free-interface normal vibration modes as component modes, as in the case of the Craig-Bampton or the Rubin method.
The normal modes accurately describe the elastodynamics within the body in the frequency band of interest, while the static deflection shapes accurately describe the local flexibility of the contact region.
%

\subsubsection*{Numerical issues with regularization}
A popular approach in the vibrations community to deal with the complementarity inequalities describing the contact constraints is penalty regularization.
This can be interpreted as replacing the rigid constraint by a unilateral spring with finite stiffness.
Important advantages of the penalty regularization are its simple implementation and that the problem degenerates into an ordinary differential equation system.
The delicate question now becomes how to properly set the penalty stiffness.
The lower it is, the more violation of the contact constraint is allowed.
For high values, on the other hand, it is well-known that the numerical treatment of the problem becomes difficult.
More specifically, an extremely fine time discretization is needed to ensure stability of explicit integration methods \cite{carp1991}, and the one-step equations of implicit methods become ill-conditioned causing numerical (nonphysical) oscillations \cite{Leine.2004,Acary.2008}.
It is therefore numerically more robust to view the contact stresses/forces as Lagrange multipliers enforcing the contact constraints and apply designated methods for their computation \cite{Acary.2008}.

\subsubsection*{Numerical oscillations associated with standard space discretization}
In the spatially continuous case, the relative contact velocity jumps to zero when a contact closes with nonzero speed.
Space discretization with standard finite elements associates a finite mass to all nodes, including those located at the contact boundary.
If a node with finite mass was stopped immediately upon contact, kinetic energy would be lost.
To prevent this nonphysical energy loss, the relative velocity must keep its magnitude, but change its sign to avoid penetration.
In contrast to the spatially continuous case, this leads to a finite impulse and causes spurious, high-frequency oscillations, both at the contact and within the flexible body, see \eg \cite{Acary.2013}.
The high-frequency oscillations are an important problem for time integration:
They limit the maximum stable time step for explicit schemes, and they cause high effort for the iterative solution of the contact problem in implicit schemes \cite{carp1991}.
To counteract the spurious oscillations, it is common practice to add numerical damping specifically to the high-frequency modes.
However, this may still lead to large computation effort and distort the wave propagation behavior \cite{Tschigg.2018}.
Dedicated constraint stabilization schemes have been proposed (\eg \cite{Acary.2013,Krause.2009,krause2012,Gear.1985}), which all come at the cost of considerable computational effort \cite{Acary.2013}.
Two approaches that counteract the spurious oscillations are further discussed in the following paragraphs, namely splitting the flexible body into a contact and a remaining region, and the massless boundary approach.
Other methods, such as Nitsche's method \cite{Chouly.2015} also lead to a well-posed semi-discrete problem, but do not reduce the spurious oscillations to the same extent.
\\
The \emph{splitting approach} was proposed by Seifried \etal for impact simulations \cite{Seifried.2010}.
The idea is to replace the contact region by a force-displacement relation derived from quasi-static assumptions, and to approximate the dynamics of the remaining region in terms of the classical modal truncation.
As force-displacement relation, the Hertzian contact law can be used for appropriate geometries.
As alternative, the force-displacement relation can be determined by interpolating the results of static finite element analyses.
The splitting approach has two major drawbacks:
First, it is strictly limited to one-dimensional frictionless problems with small contact areas.
Hence, the contact between conforming surfaces (as occurring in most joints), rolling-type motions or stick-slip interactions cannot be treated with this approach.
Second, the need to define an artificial boundary between statically treated contact region and dynamically treated remaining region leads to an inevitable compromise between the resolution of the contact stresses and the resolution of the elastodynamics within the whole body.
\\
The \emph{massless boundary approach} views the standard space discretization (which was not initially designed for impact problems) as the source of the problem.
When no inertia is associated to the nodal displacement degrees of freedom at the contact interface, the differential index of the mathematical problem reduces from 3 to 1, which substantially reduces the numerical contact oscillations \cite{Ascher.1998}.
The momentum balance restricted to the contact boundary is then a purely static sub-problem, and computationally robust methods are available to solve this along with the contact constraints \cite{Acary.2008}.
Another important benefit of this approach is that no impact law is needed for the boundary nodes, so one does not have the delicate task to set a coefficient of restitution.
The difference between the available massless boundary methods is how they get rid of the mass at the boundary.
Renard \etal \cite{Khenous.2008}, who introduced the massless boundary approach, proposed to determine the mass redistribution as the solution of an optimization problem (the objective being to shift around as little mass as possible) under the constraints of retaining the 0th, 1st and 2nd mass moment of inertia.
Besides the high computational effort of this strategy, it is rather questionable to what extent these constraints are relevant for vibration or wave propagation problems.
Wohlmuth \etal proposed to relocate the integration points away from the boundary in such a way that no mass is associated with the contact nodes \cite{Hager.2008,Hager.2009}.
Both Renard \etal \cite{Renard.2010} and Wohlmuth \etal \cite{Tkachuk.2013} later refined their methods by using shape functions different from the standard ones.

\subsubsection*{Enormous effort to compute long-term behavior (vibrations)}
As explained above, it is already challenging to simulate a short time interval around a single collision.
For vibration problems, the steady-state behavior is of primary concern.
The critical excitation occurs usually in the lowest-frequency vibration modes.
Damping is relatively light (otherwise one would not have any vibration problem).
Thus, long transients, and hence many cycles of the lowest-frequency modes must be simulated to determine the limit state.
To ensure vibration safety, many of such simulations are generally necessary for design optimization and uncertainty quantification, due to the nonlinear dependence on excitation conditions, and due to the dependence of the long-term behavior on the initial conditions (coexisting steady states).

\subsection*{Purpose and outline of the present work}
We propose to unite the massless boundary concept (originating from computational mechanics) and the idea of component mode synthesis (common in structural dynamics).
Time stepping schemes for massless boundary models are proposed in \sref{timestepping}.
Component mode synthesis methods yielding a massless boundary are addressed in \sref{masslessCMS}.
The computational performance is assessed for a series of benchmarks in \sref{results}.
This article ends with the conclusions in \sref{conclusions}.

\nc{\qq}{\mm q}
\nc{\uu}{\mm u}
\nc{\qqb}{\qq_{\mathrm b}}
\nc{\qqi}{\qq_{\mathrm i}}
\nc{\ui}{{\uu}_{\mathrm{i}}}
\nc{\dui}{\dot{\uu}_{\mathrm{i}}}
\nc{\ub}{{\uu}_{\mathrm{b}}}
\nc{\fex}{\mm f}
\nc{\fexb}{{\mm f}_{\mathrm b}}
\nc{\fexi}{{\mm f}_{\mathrm i}}
\nc{\Wb}{\mm W_{\mathrm b}}
\nc{\Wbtra}{\mm W\tra_{\mathrm b}}
\nc{\Nb}{B}
\nc{\Ni}{I}
\nc{\Nc}{C}
\nc{\half}{\frac12}
\nc{\qqbk}{\qqb^k}
\nc{\qqik}{\qqi^k}
\nc{\qqikm}{\qqi^{k-1}}
\nc{\qqikp}{\qqi^{k+1}}
\nc{\qqbkm}{\qqb^{k-1}}
\nc{\qqbkp}{\qqb^{k+1}}
\nc{\ubk}{\ub^k}
\nc{\ubkm}{\ub^{k-1}}
\nc{\uikph}{\ui^{k+\half}}
\nc{\uikmh}{\ui^{k-\half}}
\nc{\fexh}{\hat{\mm f}}
\nc{\fexbh}{\hat{\mm f}_{\mathrm b}}
\nc{\fexih}{\hat{\mm f}_{\mathrm i}}
\nc{\qtil}{\tilde{\qq}}
\nc{\Mtil}{\tilde{\mm M}}
\nc{\Ktil}{\tilde{\mm K}}
\nc{\Dtil}{\tilde{\mm D}}
\nc{\eye}{\mm I}
\nc{\nmod}{N_{\mathrm{mod}}}
\nc{\ncon}{N_{\mathrm{con}}}

\section{Time step integration of a massless boundary model\label{sec:timestepping}}
In the following, we develop a simulation procedure for a massless boundary model.
The procedure is directly applicable to a finite element model, provided that the mass has been accordingly redistributed.
We later propose component mode synthesis methods to achieve (a) the massless boundary starting from a conventional finite element model, and (b) a reduction of the number of generalized coordinates describing the inner dynamics.
\\
Consider the equations of motion of a massless boundary model subjected to contact constraints,
\ea{
	\mm K_{\mathrm{bb}}\qqb + \mm K_{\mathrm{bi}}\qqi - \Wb\mm\lambda &=& \fexb(t) \fk\label{eq:aeq}\\
	\mm M_{\mathrm{ii}}\dui + \mm D_{\mathrm{ii}}\ui + \mm K_{\mathrm{ii}}\qqi + \mm K_{\mathrm{ib}}\qqb &=& \fexi(t) \fk\label{eq:deq}\\
	\ub = \dot{\qq}_{\mathrm b}\fk \quad \ui = \dot{\qq}_{\mathrm i}\fp && \label{eq:qu}
}
Herein, $\qqb\in\mathbb R^{\Nb\times 1}$, $\qqi\in\mathbb R^{\Ni\times 1}$ are the vectors of displacements at the contact boundary, and remaining generalized coordinates, respectively, and $\Nb$ and $\Ni$ are the respective dimensions.
Overdot denotes derivative with respect to time $t$.
$\mm K_{\mathrm{bb}}\in\mathbb R^{\Nb\times\Nb}$, $\mm K_{\mathrm{bi}} = \mm K\tra_{\mathrm{ib}}\in\mathbb R^{\Nb\times\Ni}$, $\mm K_{\mathrm{ii}}\in\mathbb R^{\Ni\times\Ni}$ are the respective partitions of the symmetric and positive semi-definite stiffness matrix.
$\mm M_{\mathrm{ii}}\in\mathbb R^{\Ni\times\Ni}$ is the symmetric and positive definite mass matrix associated with $\qqi$.
Viscous damping is considered within the bodies (matrix $\mm D_{\mathrm{ii}}\in\mathbb R^{\Ni\times\Ni}$), but not at the boundary\footnote{Note that the damping associated with the contact boundary will in most cases be dominated by dry friction rather than viscous damping, so that this does not seem to be a strong limitation.}.
$\fexb\in\mathbb R^{\Nb\times 1}$, $\fexi\in\mathbb R^{\Ni\times 1}$ denote the imposed forcing with known explicit time dependence.
$\mm\lambda\in\mathbb R^{\Nc\times 1}$ is the vector of $\Nc$ Lagrange multipliers which can be interpreted as contact forces, with the constant matrix of contact force directions $\Wb\in\mathbb R^{\Nb\times \Nc}$.
The algebraic equations \erefo{aeq} describe the quasi-static force balance at the boundary.
The ordinary differential equations \erefo{deq} describe the dynamic generalized force balance within the body.
We limit the development to linear elasticity and linear kinematics; the extension of the method to, \eg, geometric and material nonlinearity, or nonlinear and time-variant contact kinematics (with $\Wb(\qq,t)$ rather than constant $\Wb$) is left for future work.
Concerning geometric nonlinearity described by multivariate polynomials, such an extension could rely on the Stiffness Evaluation Procedure or Implicit Condensation \cite{Hollkamp.2008}, which can be combined with Component Mode Synthesis \cite{Kuether.2015b}.
Concerning other nonlinearities, such an extension could rely on a form of hyper-reduction.
\\
The contact gap $\mm g$ and gap velocity $\mm \gamma$ are kinematically related to $\qqb$, $\ub$ via \eref{ckinematics},
\ea{
	\mm g = \Wbtra \qqb + \mm g_0(t) \quad \mm\gamma = \Wbtra \ub + \dot{\mm g}_0(t) \fk \label{eq:ckinematics}\\
	-\mm\gamma \in \mathcal N_{\mathcal C}\left(\mm\lambda\right) \fp  \label{eq:claws}
}
The contact laws in \eref{claws} are here expressed as inclusion into the normal cone, $\mathcal N_{\mathcal C}$, to the admissible set, $\mathcal C$, of the contact forces $\mm\lambda$.
They are formulated on velocity level here and apply to all active contacts (closed normal gap).
Unilateral interaction (Signorini conditions) is considered in the normal contact direction.
Coulomb's law of dry friction is considered in the tangential contact plane.
With $\mm\gamma$ set up as defined in \eref{gammasetup}, the set $\mathcal C$ is given by \eref{admissibleset},
\ea{
	\mm\gamma = \vector{\mm\gamma_1 \\ \vdots \\ \mm\gamma_{\ncon}}\fk \quad \mm\gamma_j = \vector{\gamma_{\mathrm n,j} \\ \mm\gamma_{\mathrm t,j}} \fk \label{eq:gammasetup}\\
	\mathcal C = \mathcal C_1 \times \ldots \times \mathcal C_{\ncon}\fk \quad \mathcal C_j = \mathbb R_0^+ \times \mathcal D\left(\mu\lambda_{\mathrm n,j}+\mu\lambda_{\mathrm n,j}^0\right)\fp \label{eq:admissibleset}
}
Herein, $\mathcal D\left(r\right)$ denotes the planar disk of radius $r$.
$\mm g$ and $\mm\lambda$ are set up analogous to \eref{gammasetup}.
In the case of a preloaded contact point, $\lambda_{\mathrm n,j}^0>0$ is the normal preload; $\lambda_{\mathrm n,j}^0=0$ for initially open contacts.
\\
The proposed approach is able to deal with contact among multiple flexible bodies, self-contact, and contact with a rigid wall.
Also, the above formulations are applicable regardless of the contact discretization (\eg node-to-segment, segment-to-segment/Mortar).
The $\ncon$ contact gaps generally correspond to certain integration points.
The contact forces, $\mm\lambda$, are generally obtained by numerical integration of the contact stress field.
The contact force at a certain integration point is simply the contact stress times the associated area (weight of numerical integration), see \eg \cite{Krack.2016}.
The numerical examples are limited to contact with the (rigid) ground and, for simplicity, we use a node-based integration with equal weights (rather than a consistent integration).
In the remainder of this paper, thus, the distinction between nodes and integration points, and the case of non-matching nodes do not receive any particular attention.
%

\subsection{Time stepping schemes}
A variety of integrators can be applied to the constrained differential-algebraic equation systems.
A comparison of several integrators can be found in \cite{Doyen.2011,Dabaghi.2014}.
In the course of the present work, various integrators were tested as well.
In particular, we tried higher-order schemes, but found that their performance suffered from numerical oscillations again.
Based on our experience, we propose the relatively simple Verlet scheme in its leapfrog form.
It should be remarked that the Moreau-like integrators, which are very popular in the mass-carrying case, rely on the same scheme, but are usually formulated in the kick-drift-kick form rather than in the leapfrog form.
As the symmetric Moreau-like integrator developed in \cite{Capobianco.2018}, Verlet's scheme enjoys time-reversibility.
Moreover, it has favorable energy conservation properties thanks to its symplectic nature.
Under fixed contact conditions, the approach enjoys second-order accuracy.
Verlet's scheme leads to an explicit integration scheme for the inner coordinates, while the static contact problem restricted to the boundary is solved implicitly.
Consequently, the scheme is only conditionally stable.
The details of the proposed time stepping scheme are described in the following.
\\
In the leapfrog form, the grids of velocities and coordinates are shifted.
The time levels for $\uu$ are in the middle of the time levels for $\qq$.
We use the notation $\qq^k=\qq(t^k)$, where $t^k$ is the $k$-th time level, and $\uu^{k+\half}=\uu(t^k+\frac{\Delta t}2)$ with the time step $\Delta t$.
Then Verlet's scheme can be expressed as
\ea{
	\qq^{k+1} &=& \qq^{k}+\uu^{k+\half}\Delta t \fk \label{eq:qtrapz}\\
	\uu^{k+\half} &=& \uu^{k-\half} + \dot{\uu}^{k}\Delta t \fp \label{eq:utrapz}
}
For a function $\mm h\left(\qq,\uu,t\right)$, one can use the approximation
\e{
	\mm h^k = \frac12 \left(~\mm h\left(\mm q^k,\mm u^{k-\half},t^k\right) + \mm h\left(\mm q^k,\mm u^{k+\half},t^k\right)~\right)\fk \label{eq:htrapz}
}
and in particular, $\uu^k = \frac12\left(\uu^{k-\half}+\uu^{k+\half}\right)$, $\qq^{k-\half} = \frac12\left(\qq^{k-1}+\qq^k\right)$.
\\
In the following, it is useful to treat frictionless and frictional contact separately.
The implementation of the leapfrog scheme for the given problem setting is illustrated in \fref{TimeLines}.

\subsubsection{Frictionless contact}
In the case of frictionless contact, the contact constraints can be formulated on the displacement level.
Hence, the contact constraints in \eref{claws} can be replaced by $-\mm g \in \mathcal N_{\mathcal C}\left(\mm\lambda\right)$.
The proposed algorithm is:
\begin{enumerate}
	\item Set $k=1$, $t^1=t_{\mathrm{start}}$ and initial $\qqi^1$, $\ui^{\half}$.
	\item Solve \eref{aeq}, \eref{ckinematics} and the contact constraints on \emph{displacement level} evaluated at $t^k$ with respect to $\qqbk$, $\mm\lambda^k$.
	\item Solve \eref{deq} evaluated at $t^k$ with respect to $\uikph$.
	\item Evaluate \eref{qtrapz} to determine $\qqi^{k+1}$.
	\item If $t^k<t_{\mathrm{end}}$, increase $k$ by one and go back to step 2.
\end{enumerate}
Note that the boundary velocities $\ub$ are not needed in this algorithm, and thus no stepping is applied to the boundary coordinates.

\subsubsection{Frictional contact}
In this case, the contact constraints cannot be formulated on the displacement level, and are instead formulated on the velocity level.
Hence, the boundary velocities $\ub$ are needed and a stepping rule must be employed to relate the time discrete forms of $\ub$ and $\qqb$.
As in conventional Moreau-like schemes, this stepping rule for $\qqb$ is then substituted into \eref{aeq} to obtain an expression for the contact gap velocities as function of the contact forces.
%
%
The proposed algorithm is:
\begin{enumerate}
	\item Set $k=1$, $t^1=t_{\mathrm{start}}$ and initial $\qqi^1$, $\ui^{\half}$, $\qqb^0$.
	\item Solve \eref{aeq}, \eref{ckinematics} and the contact constraints on \emph{velocity level} evaluated at $t^k$ with respect to $\ub^{k-\half}$, $\mm\lambda^{k}$, and update $\qqbk$ using \eref{qtrapz}.
	\item Solve \eref{deq} evaluated at $t^k$ with respect to $\uikph$.
	\item Evaluate \eref{qtrapz} to determine $\qqi^{k+1}$.
	\item If $t^k<t_{\mathrm{end}}$, increase $k$ by one and go back to step 2.
\end{enumerate}
Note that steps 3-5 are identical to the frictionless case.
As opposed to the frictionless case, an initial value for $\qqb^0$ is needed in the frictional case.
\begin{figure}[htb]
	\centering
	\includegraphics[width=0.99\textwidth]{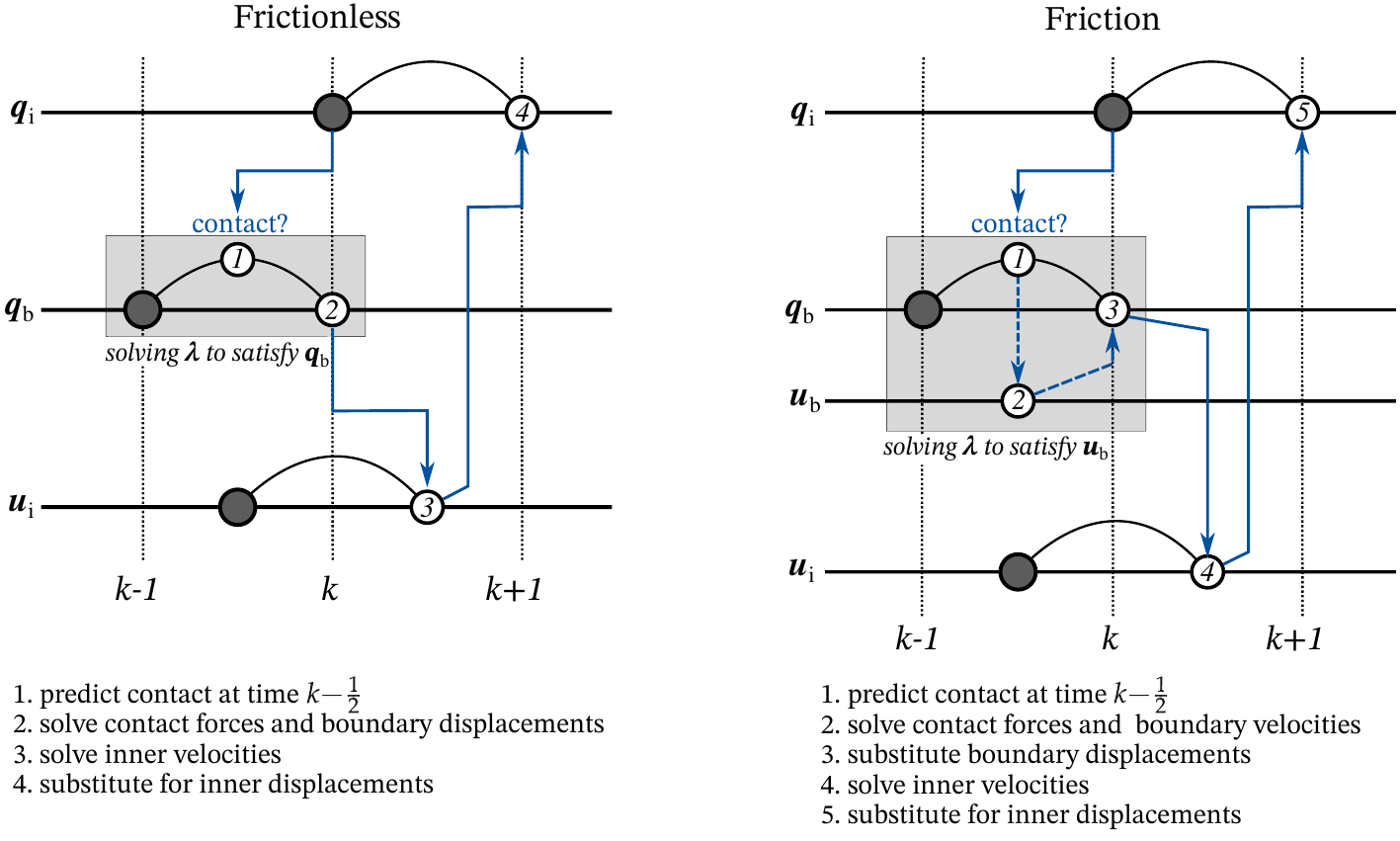}
	\caption{Schematic illustration of the time integration schemes}
	\label{fig:TimeLines}
\end{figure}

\subsection{Active set strategy}
The contact constraints are treated using an active set strategy; \ie, the contact problem is restricted to the active set of constraints.
To determine the active set of constraints, $\mathcal I_{\mathrm a}$, two variants are proposed depending on whether a particular contact interface is initially open or preloaded.
\\
In the \emph{initially open case}, the normal gaps are predicted with \eref{aeq} assuming $\mm\lambda=\mm 0$:
\nc{\Kbb}{\mm K_{\mathrm {bb}}}
\nc{\Kbi}{\mm K_{\mathrm {bi}}}
\ea{
	\mm g^{\mathrm{pre}} &=& \mm g_0\left(t^k\right) + \Wbtra\Kbb^{-1}\left(~\fexb\left(t^k\right) ~-~ \Kbi \qqik~\right)\fk \\
	\mathcal I_{\mathrm a} &=& \lbrace j\vert g_{\mathrm n,j}^{\mathrm{pre}} \leq 0 \rbrace\fp
}
The active constraints are those with non-positive normal gap.
\\
In the \emph{preloaded case}, the contact forces are predicted with \eref{aeq} assuming $\qqbk=\qqbkm$ (sticking contact).
\ea{
\mm\lambda^{\mathrm{pre}} &=& \Wb^{-1}\left(~\Kbb\qqbkm~+~\Kbi\qqik~-~\fexb\left(t^k\right)~\right)\fk \label{eq:lambdapre}\\
\mathcal I_{\mathrm a} &=& \lbrace j\vert \lambda_{\mathrm n,j}^{\mathrm{pre}}+\lambda_{\mathrm n,j}^{0} \leq 0 ~ \lor ~ \|\mm\lambda_{\mathrm t,j}^{\mathrm{pre}}\| \geq \mu \lambda_{\mathrm n,j}^{\mathrm{pre}}+\mu \lambda_{\mathrm n,j}^{0} \rbrace\fp
}
The active constraints are those with non-positive normal contact forces, and/or tangential forces not inside the Coulomb cone in the frictional case.
To solve \eref{lambdapre}, $\Wb$ must be invertible, which is the case under the assumptions outlined in \sref{masslessCMS}.
\fig[ph!]{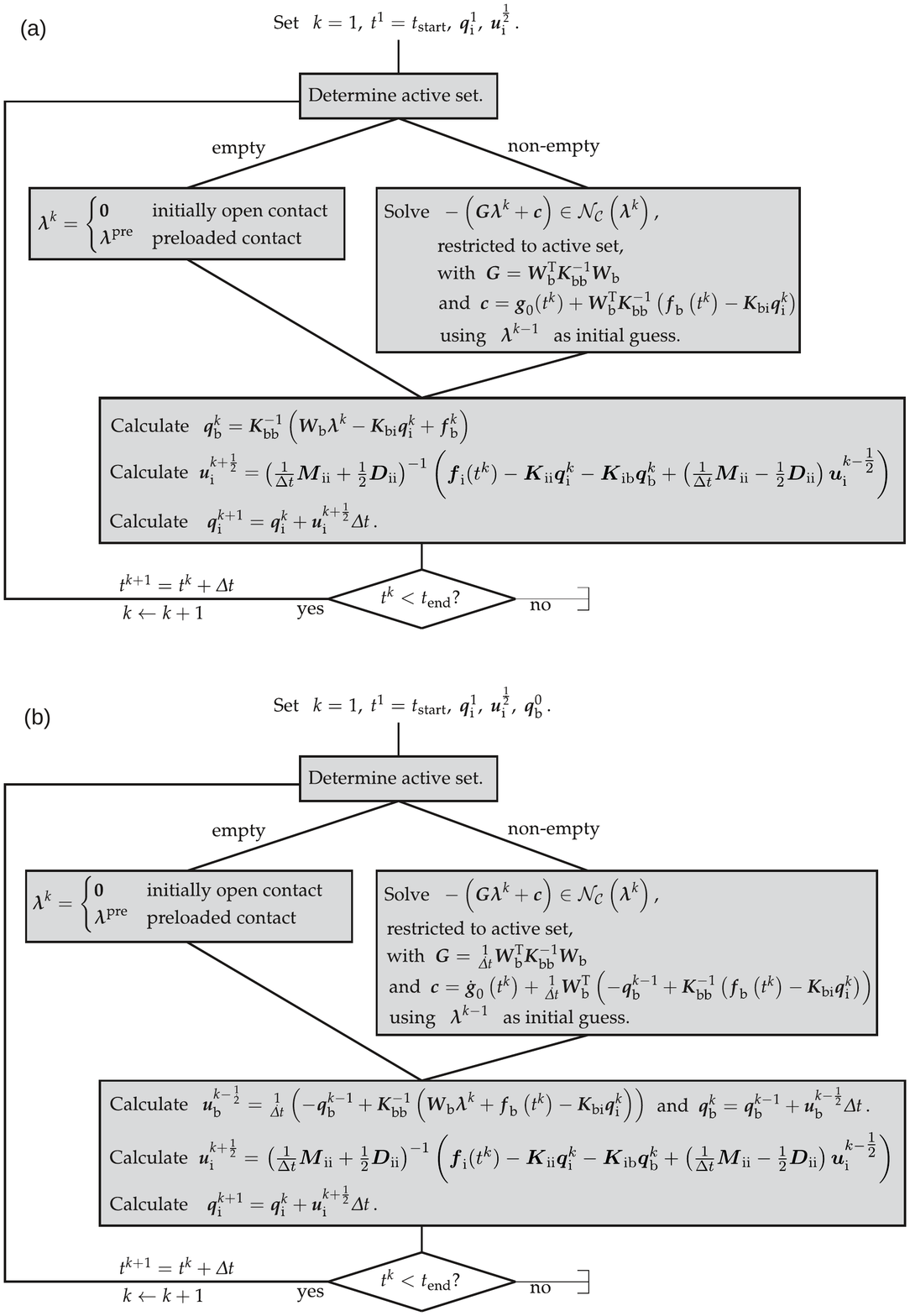}{Overview of the simulation algorithms for the cases of (a) frictionless contact, (b) frictional contact}{0.9}
\\
The important advantage of the described procedure is the non-iterative solution of the contact problem if the initially open contact remains open and the preloaded contact remains closed (or sticking in the frictional case).
If a system contains both initially open and preloaded interfaces, the described procedure can be easily applied separately for each interface, provided that the interfaces are \emph{elastically and kinematically decoupled} in the sense that the respective coupling partitions of $\mm K_{\mathrm{bb}}$ and $\Wb$ are zero.
An alternative to the described procedure is to use a stepping rule, \ie, to predict $\qqb$ at the next time level using the current value of $\ub$, and then to define the constraints with non-positive normal gap as active.
Compared to the proposed procedure, we found that this leads to more iterations in the case of preloaded contact interfaces.

\subsection{Initialization}
It should be remarked that Moreau-like schemes commonly use the notation of starting, mid- and end point of a time interval (kick-drift-kick form).
In the leapfrog form used above, actually $t^k$ corresponds to the midpoint, $t^{k-\half}$ and $t^{k+\half}$ are starting and end point, respectively.
Note that stepping from midpoint to midpoint of the coordinates $\qq$ is possible since the corresponding velocity is assumed as constant in Moreau-like schemes.
The proposed formulation involving a shifted grid of $\qq$ and $\uu$ is, of course, fully equivalent.
However, a caveat is that the initialization is slightly more complicated if values of $\qq$ and $\uu$ are imposed at the same time instant $t_{\mathrm{start}}$.
This can be addressed properly by implementing an initialization step as proposed in \cite{Capobianco.2018}.
As alternative, one may omit this initialization and simply use the approximation $\ui^{\half} = \ui(t^1)$, and $\qqb^0 = \qqb(t^1)$ in the frictional case.
This should be valid if the initial velocity changes only slowly in the beginning of the simulation. 
To reduce the error introduced by this approximation, one could also use smaller time steps at the beginning of the simulation.
It should be remarked that in many cases, the long-term behavior is of primary interest and the dependence on the exact initial conditions plays only a minor role.
It is also useful to note that the required initial data is readily available if the simulation is resumed using the same algorithm.

\subsection{Summary of the simulation algorithms}
The simulation algorithms for the frictionless and frictional case are summarized in \fref{signalFlowPlan}.
Recall that an active set strategy is used; \ie, the terms $\Wb$, $\mm g$, $\mm \lambda$ \etc are restricted to the active set.
The global implicit algebraic inclusion problem in the contact forces is solved using an augmented Lagrangian approach, as described in \ref{asec:AL}.

%

\section{Massless boundary Component Mode Synthesis\label{sec:masslessCMS}}
The algorithms proposed in \sref{timestepping} can be applied directly to a finite element model, provided that the mass has been accordingly redistributed.
Appropriate methods for this have been mentioned in the introduction.
We propose a convenient alternative, namely to achieve a massless boundary during the construction of the reduced-order model using appropriate component mode synthesis.
This has the important benefit that standard finite elements can be used for the initial model.
Also, component mode synthesis is capable of reducing the computational effort of vibro-impact processes substantially, without significantly sacrificing the accuracy, as discussed further in \sref{discuss}.
We consider both a free-interface (MacNeal) and a fixed-interface (Craig-Bampton) component mode synthesis technique.
It is expected that the former is better suited for initially open contacts, while the latter is better suited for preloaded contacts.
The Craig-Bampton method is by far the most popular Component Mode Synthesis method \cite{klerk2008}.
The MacNeal and the Rubin method also addressed in this work are by far the most popular free-interface techniques\footnote{The exact dual of the Craig-Bampton method (in a linear framework!) would rely on a substructure coupling via forces rather than displacements \cite{rixe2004}, which is not applicable to the proposed nonlinear contact algorithm.}.
Both methods rely on splitting the vector of coordinates $\qq$ into boundary coordinates $\qqb$ and inner coordinates $\qqi$,

\e{
	\qq = \vector{\qqb\\ \qqi} \simeq \mm R \vector{\qqb \\ \mm\eta} = \mm R \qtil \fk \label{eq:partitioning}
}
and approximating these in terms of component modes (columns of matrix $\mm R$) and associated coordinates.
For the ease of contact treatment, the boundary coordinates $\qqb$ are retained, while the inner coordinates are replaced by a reduced set of generalized coordinates $\mm\eta$.
All Component Mode Synthesis methods discussed in this paper (standard and massless Craig-Bampton, Rubin, MacNeal) have in common that they capture the static flexibility with respect to loads applied at the boundary in an exact way.
\\
In the case of linear contact kinematics, it can be shown that a regular transform to relative coordinates, \ie, the contact gaps, is always possible.
With appropriate sorting, the matrix $\Wb$ in \eref{aeq} and \eref{ckinematics} is then the identity matrix.
It is proposed to carry out such a coordinate transform prior to applying component mode synthesis, as this is known to improve the modal convergence.
The global problem is then mathematically equivalent to that of a single body with contact to a rigid foundation.
It is thus sufficient to discuss the method in terms of a model described by a single pair of stiffness and mass matrices, $\mm K$ and $\mm M$, partitioned in accordance with \eref{partitioning}.

\subsection{MacNeal method}
MacNeal's hybrid synthesis method is a free-interface technique \cite{macn1971}.
It is well-known that the MacNeal method yields a singular mass matrix, where no inertia is associated with the boundary coordinates.
To the authors' knowledge, however, the potential benefits of this property for the simulation of vibro-impact processes are still unknown.
For completeness, the MacNeal method is briefly described in the following.
\\
In the MacNeal method, the component modes are a subset of free-interface normal modes and residual flexibility attachment modes.
The free interface normal modes $\mm\phi_k$ with associated natural frequency $\omega_k$ are determined from the eigenvalue problem
\e{
	\left(\mm K - \omega_k^2\mm M \right)\mm\phi_k =
	\mm 0\fk \quad \mm\phi\tra_k\mm M\mm\phi_k = 1\fk
}

and are assumed to be normalized with respect to the mass matrix $\mm M$.
The set of $\nmod$ lowest-frequency modes, $\mm \Phi = \left[\mm\phi_1,\ldots,\mm\phi_{\nmod}\right]$ is retained.
The retained normal modes should cover the relevant frequency range of the response.
In the linear case, the frequency range of the response corresponds to that of the excitation, so that the modal truncation is usually a straight-forward task.
In the nonlinear case, the nonlinear forces generally generate higher frequencies so that the cutoff frequency might have to be much higher than the highest relevant excitation frequency.
\\
The flexibility matrix is denoted as $\mm F$ and defined as the inverse of the stiffness matrix, $\mm F = \mm K\inv$.
The $j$-th column of $\mm F$ represents the static deflection due to a unit load applied at the $j$-th element of $\qq$.
The columns of $\mm F$ corresponding to all boundary coordinates $\qqb$ are determined.
The associated upper and lower sub-matrices are denoted as $\mm F_{\mathrm{bb}}$ and $\mm F_{\mathrm{ib}}$, respectively.
The corresponding residual flexibility sub-matrices can then be expressed as
\ea{
	\vector{ \mm F_{\mathrm{bb}}^\prime \\ \mm F_{\mathrm{ib}}^\prime } =
	\vector{ \mm F_{\mathrm{bb}} \\ \mm F_{\mathrm{ib}} }
	- \mm\Phi \diag\left(\frac{1}{\omega_k^2}\right)\mm\Phi\tra_{\mathrm b} \fp
}
$\mm\Phi_{\mathrm b}$ is the upper part of matrix $\mm \Phi$ associated with the boundary $\qqb$, while $\mm\Phi_{\mathrm i}$ is the lower part associated with the inner coordinates $\qqi$.
\\
The matrix of component modes is
\e{ \mm R = \matrix{cc}{
		\eye                                                                 & \mm 0 \\
		\mm F_{\mathrm{ib}}^\prime\left(\mm F_{\mathrm{bb}}^\prime\right)\inv &
		\mm\Phi_{\mathrm i} - \mm F_{\mathrm{ib}}^\prime\left(\mm F_{\mathrm{bb}}^\prime\right)\inv\mm\Phi_{\mathrm b}
	}\fp
}
Herein, $\eye$ denotes the identity matrix (of proper dimension), $\mm F_{\mathrm{bb}}^\prime$ and $\mm F_{\mathrm{ib}}^\prime$ are the corresponding partitions of the residual flexibility matrix, and $\mm\Phi_{\mathrm b}$ and $\mm\Phi_{\mathrm i}$ are the corresponding partitions of the matrix containing the truncated set of free-boundary mode shapes as columns.
The MacNeal method relies on the following reduced mass and stiffness matrices:
\ea{
	\Mtil = \matrix{cc}{\mm 0 & \mm 0 \\ \mathrm{sym.} & \eye}\fk \quad
	\Ktil = \matrix{cc}{\left(\mm F_{\mathrm{bb}}^\prime\right)\inv & - \left(\mm F_{\mathrm{bb}}^\prime\right)\inv \mm\Phi_{\mathrm b} \\ \mathrm{sym.} & \diag\left(\omega_k^2\right) + \mm\Phi\tra_{\mathrm b} \left(\mm F_{\mathrm{bb}}^\prime\right)\inv\mm\Phi_{\mathrm b}}\fp \label{eq:MacNeal}
}
Apparently, the MacNeal method readily yields a massless boundary model, with a very simple form of the mass matrix, which is amenable to the simulation procedures developed in \sref{timestepping}.
\\
It should be noted that $\Ktil = \mm R\tra \mm K\mm R$.
Using $\Mtil = \mm R\tra \mm M\mm R$ instead of the relation in \eref{MacNeal} yields the \emph{Rubin method}.
In the case of the Rubin method, $\Mtil$ is positive definite, which corresponds to a conventional, mass-carrying boundary model.
The Rubin method serves as a reference in one of the numerical examples.
Defining the reduced matrices as quadratic product with $\mm R$ can be shown to be consistent with the Galerkin requirement that the error made by the approximation $\qq=\mm R\qtil$ should be orthogonal with respect to the component modes.
Hence, the Rubin method is a Galerkin method, in contrast to the MacNeal method, which is why the MacNeal method is sometimes called \emph{inconsistent}.
In the MacNeal method, the mass that should be carried by the residual flexibility attachment modes is neglected.
Consequently, the MacNeal method describes the dynamic stiffness less accurately than the Rubin method, particularly if only a small number of modes, $\nmod$, is retained.
On the other hand, the mass-deficiency with respect to the boundary makes the MacNeal method attractive for dynamic contact problems.
\\
If rigid body motion is possible, the stiffness matrix $\mm K$ is singular, and the MacNeal method cannot be applied directly.
One means to overcome this is to prevent rigid body motion by adding some artificial stiffness \eg to the boundary, apply the MacNeal method, and subtract the same stiffness term again after the reduction.

\subsection{Massless Craig-Bampton method}
As opposed to the MacNeal method, the standard Craig-Bampton method does not yield a massless boundary \cite{crai1968}.
Indeed, if the parent model has a positive definite mass matrix, as in the case of a standard finite element model, the Craig-Bampton method preserves the positive definiteness.
In the following, we briefly recap the standard Craig-Bampton method and suggest adjustments to enforce a massless boundary.
A similar procedure has been proposed in \cite{sher2013}.
\\
In the Craig-Bampton method, the matrix of component modes is
\ea{
	\mm R = \matrix{cc}{
		\eye     & \mm 0 \\
		\mm \Psi &
		\mm\Theta}\fk
	\label{eq:CB}
}
where $\mm \Theta$ denotes the set of $\nmod$ lowest-frequency fixed-interface normal modes (restricted to the inner coordinates), and the first columns correspond to the static constraint modes with $\mm\Psi = -\mm K\inv_{\mathrm{ii}}\mm K_{\mathrm{ib}}$.
Concerning the modal truncation, what was stated about the MacNeal method still applies.
The $j$-th column of $\mm\Psi$ represents the static deflection of $\qqi$ due to unit displacement imposed at the $j$-th element of $\qqb$ (with the other boundary coordinates fixed).
The reduced mass and stiffness matrices satisfy $\Mtil = \mm R\tra\mm M\mm R$ and $\Ktil = \mm R\tra\mm K\mm R$, and can be expressed as
\ea{
	\Mtil = \matrix{cc}{\Mtil_{\mathrm{bb}} & \Mtil_{\mathrm{bi}} \\ \mathrm{sym.} & \eye} \fk \quad
	\Ktil = \matrix{cc}{\Ktil_{\mathrm{bb}} & \mm 0 \\ \mathrm{sym.} & \diag\left(\omega_k^2\right)} \fk \label{eq:CBred}
}
where $\omega_k$ are now the natural frequencies associated with the fixed-interface normal modes.
In general, $\Mtil_{\mathrm{bb}}\neq \mm 0$ and $\Mtil_{\mathrm{bi}}\neq \mm 0$, so that inertia forces are associated with the boundary coordinates.
\\
Now, the massless boundary is achieved in two steps:
First, boundary and inner coordinates are decoupled with respect to inertia forces by a suitable regular coordinate transform.
Second, the mass associated with the boundary coordinates is removed.
For the first step, a linear combination of the fixed-interface normal modes is added to the static constraint modes,
\ea{
	\mm R_\alpha = \matrix{cc}{
		\eye                        & \mm 0 \\
		\mm \Psi-\mm\Theta\mm\alpha &
		\mm\Theta} \fp
}
The coefficient matrix $\mm\alpha$ is determined by the requirement that $\Mtil_{\mathrm {bi}}$ vanishes when replacing $\mm R$ by $\mm R_\alpha$.
This leads to
\e{
	\mm\alpha =\mm\Theta\tra\left( \mm M_{\mathrm{ib}} + \mm M_{\mathrm{ii}}\mm\Psi\right)\fp
}
This, in turn, leads to the reduced mass and stiffness matrices
\ea{
	\Mtil = \matrix{cc}{\Mtil_{\mathrm{bb}}^* & \mm 0 \\ \mathrm{sym.} & \eye} \approx \matrix{cc}{\mm 0 & \mm 0 \\ \mathrm{sym.} & \eye} \fk \quad
	\Ktil = \matrix{cc}{\Ktil_{\mathrm{bb}}^* & \Ktil_{\mathrm {bi}}^* \\ \mathrm{sym.} & \diag\left(\omega_k^2\right)} \fp \label{eq:CBdec}
}
Thanks to the linear independence of the component modes, the reduced model in \eref{CBdec} is fully equivalent to that of the standard Craig-Bampton method.
Apparently, boundary and inner coordinates are now decoupled with respect to inertia forces, but an elastic coupling is now present.
The exact opposite is the case in the standard Craig-Bampton method (\eref{CBred}).
\\
The second step now consists in simply neglecting the inertia forces associated with the new set of boundary coordinates by setting $\Mtil_{\mathrm{bb}}^*=\mm 0$.
The resulting reduced mass matrix is identical to that obtained by the MacNeal method (\eref{MacNeal}).
As the inconsistency is introduced only in the inertia terms, the static flexibility with respect to the boundary is still exactly captured.
Moreover, the fixed-interface normal modes are still exactly represented.
However, the free-interface dynamics is less accurately modeled, and can be expected to converge more slowly with the number of retained fixed-interface modes, as compared with the standard Craig-Bampton method.

\subsection{Discussion of model order reduction and computational speed-up}\label{sec:discuss}
It is useful to discuss the model order reduction achieved for typical vibro-impact processes.
State-of-the-art finite element models of single-body problems often comprise a number of $10^5$–$10^7$ nodal degrees of freedom.
Such a fine spatial discretization is often needed to accurately resolve the oscillatory stresses at critical locations and to predict the effect of intricate geometrical features on the higher-frequency normal modes.
Assuming that the body is bulky and the contact area has a simple geometry, or the body is slender and the contact area is small compared to the overall surface, the contact interface comprises typically $10^2$–$10^3$ number of nodal degrees of freedom.
The mathematical model order obtained using component mode synthesis as described in \sref{masslessCMS} (MacNeal or (massless) Craig-Bampton method) equals the number of degrees of freedom at the contact boundary, plus a number of retained normal vibration modes.
Modes up to a reasonably high frequency need to be retained to accurately resolve the wave propagation.
The highest natural frequency increases approximately linearly with the modal truncation order.
As shown in \cite{Seifried.2010}, less than $10^2$ normal modes are sufficient in most cases to accurately resolve the elastodynamics following an impact event.
With this, the model order is reduced by 2 to 5 orders of magnitude.
As explained, this goes along with a substantial reduction of the highest relevant frequency, so that larger time steps are appropriate in the numerical integration.
In this way, the smaller model order leads to reduced computational burden both by allowing larger time steps and by decreasing the effort of the linear algebra operations.
It must be emphasized, however, that sparsity is generally lost by the proposed model order reduction techniques.
That is why the computational speed-up will not simply be the ratio of the model orders.

\section{Similarities and differences to the quasi-static treatment of the high-frequency modes}
It has been proposed by other researchers to improve the simulation of impact problems by treating the high-frequency modes quasi-statically \cite{Tschigg.2018,Sherif.2012}.
Hence, the inertia associated to the high-frequency modes is neglected.
This is somewhat similar to the idea proposed in the present work.
An important benefit of the proposed approach is that no empirical criterion is needed to divide the dynamics into high- and low-frequency part, but this division follows immediately from the definition of the contact boundary and the finite element discretization.
Also, the proposed approach does not require to solve the eigenvalue problem of the reduced system (to sort the modes by frequency).

\section{Numerical results\label{sec:results}}
The computational performance of the approach developed in \sref{timestepping} is now assessed for a series of benchmarks:
(1) a one-dimensional bar dropped on a rigid ground, (2) a harmonically driven three-dimensional plate under frictionless contact, and (3) a rotating blade undergoing frictional impacts with an oval casing.
The proposed approach is compared against state-of-the-art methods.
For each benchmark, we also compared MacNeal method with the massless Craig-Bampton method.
In full agreement with other studies in linear \cite{klerk2008} and nonlinear \cite{bata2007} structural dynamics, we found that both methods overall perform similarly well.
The free-interface (MacNeal) method usually performs slightly better for initially open contacts, whereas the fixed-interface (massless Craig-Bampton) method performs slightly better for preloaded contacts.
The results of the direct comparisons are not show for brevity.

\subsection{One-dimensional bar dropped on rigid ground}
\label{sub:unidimensional_bar}
The bouncing bar is a common benchmark to assess computational methods for elastodynamic contact problems \cite{Doyen.2011,Dabaghi.2014,Dabaghi.2016,Schreyer2016,Dabaghi.2019}.
An important benefit of this benchmark is that an exact solution of the time- and space-continuous problem is available \cite{Doyen.2011}.
The problem setting is illustrated in \fref{ex1_fig_pogo}.
The bar with density $\rho$, Young's modulus $E$ and length $\ell$ is released, without initial deformation or velocity, from height $q_0$, under gravity (imposed acceleration $a_g>0$)
The displacement $q(x,t)$ of the bar is governed by the equations 
\ea{
	\rho \left(\ddot q+a_g\right) - E\frac{\partial^2 q}{\partial x^2} - \lambda = 0 \quad x\in [0,\ell]\fk\,\, t\in \mathbb R^+\fk \label{eq:barEoM}\\
	q\left(0,t\right) \geq 0\fk \quad \lambda \geq 0\fk \quad \lambda~q\left(0,t\right) = 0 \fk\label{eq:barCon} \\
	\frac{\partial q}{\partial x}\left(\ell,t\right) = 0 \fk \label{eq:barBC} \\
	q(x,0) = q_0\fk \quad \dot q(x,0) = 0 \fp \label{eq:barIC}
}
Herein, overdot denotes derivative with respect to time $t$.
The spatial coordinate $x$ refers to the undeformed configuration.
\eref{barEoM} is a one-dimensional wave equation. 
\erefs{barCon} describe the unilateral contact between the bar's lower end and the rigid ground, where the Lagrange multiplier $\lambda$ can be interpreted as normal contact pressure (force per unit cross section).
\eref{barBC} describes the free boundary condition at the bar's upper end.
\erefs{barIC} specify the initial conditions at time $t=0$.
\fig[t]{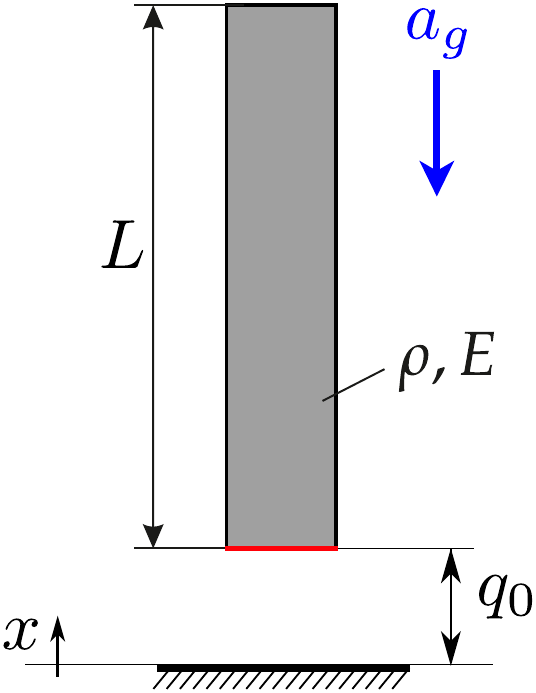}{One-dimensional bar dropped on rigid ground}{0.2}
\fig[th!]{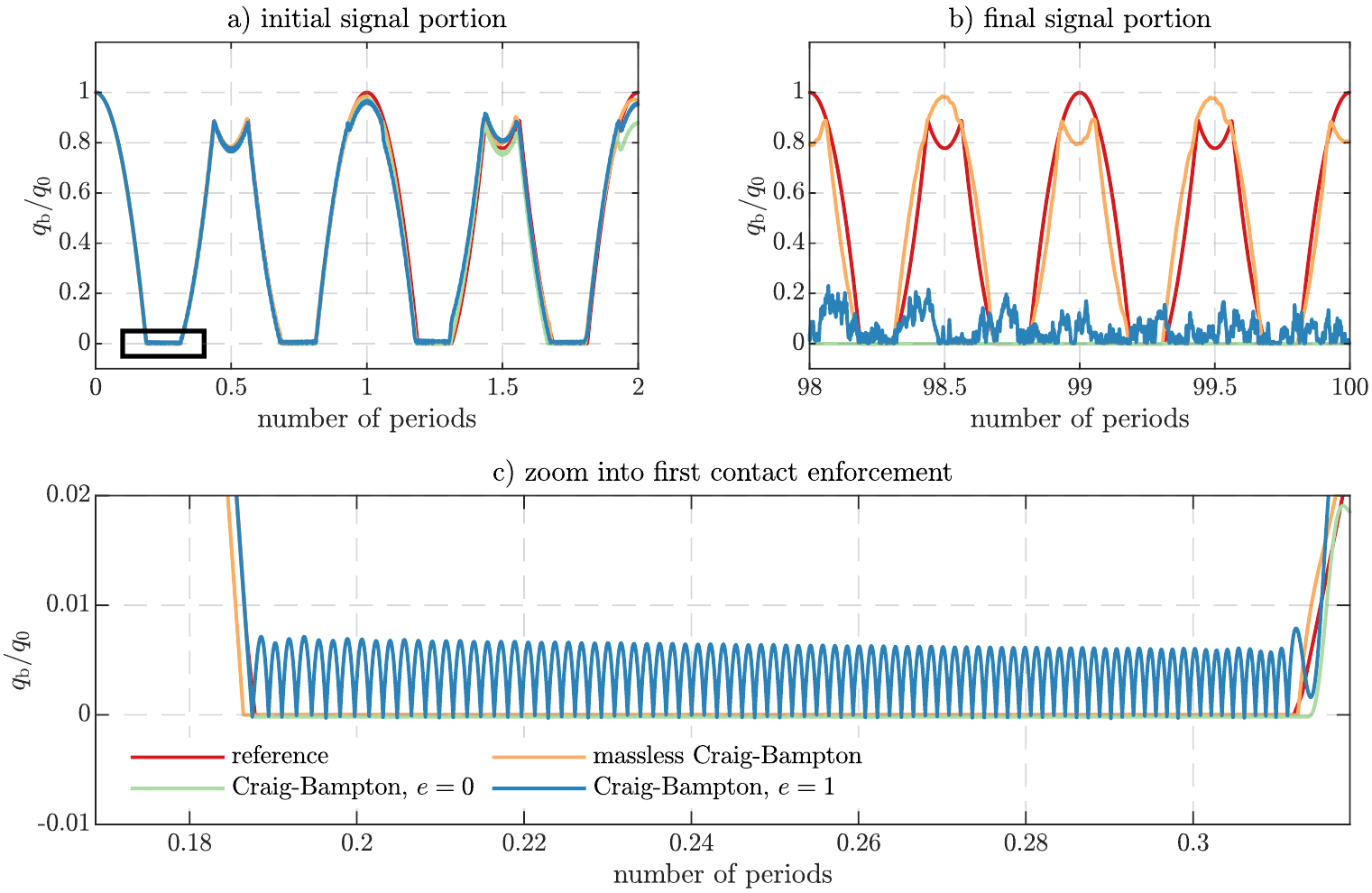}{Displacement $q_{\mathrm b}$ of the bar's lower boundary: (a) initial time range, (b) later time range, (c) zoom into first contact phase}{1.0}
\\
The parameters are specified as $\rho=1$, $E=900$, $\ell=10$.
Upon release, the bar falls freely until it makes contact with the ground.
The contact occurs with a finite duration, and initiates a shock wave which propagates with the speed $c= \sqrt{E/\rho} = 30$ through the bar.
At a certain time, the contact is released and the bar lifts up again.
The resulting displacement during the free-flight phase is a superposition of a rigid body motion with parabolic time-dependence and elastic vibrations.
The initial height and gravity acceleration were selected such that this bouncing behavior is time-periodic \cite{Doyen.2011}.
For the numerical solution, we first discretize the problem domain with a uniform mesh of finite elements with linear shape functions (element length $\Delta x=10^{-2}$).
We then apply the massless and the standard Craig-Bampton method to obtain a reduced massless and mass-carrying boundary model, respectively, using a single static constraint mode associated with the boundary coordinate $q_{\mathrm b}(t)=q(0,t)$ and retaining the $20$ lowest-frequency fixed-interface normal modes.
The simulation is carried out using the time step integration schemes described in \sref{timestepping} and \ref{asec:SymMoreau}, respectively.
In the mass-carrying case, a coefficient of restitution needs to be specified in accordance with Newton's impact law.
For the one-dimensional problem here, this coefficient determines the ratio between post- and pre-impact normal velocities of the boundary node.
We show results for the extreme cases of $\CoR=0$ and $\CoR=1$.
A time step of $\Delta t = 10^{-4}$ is used for both time stepping schemes, corresponding to a Courant number of $c\Delta t/\Delta x=0.3$
\\
\fig[t]{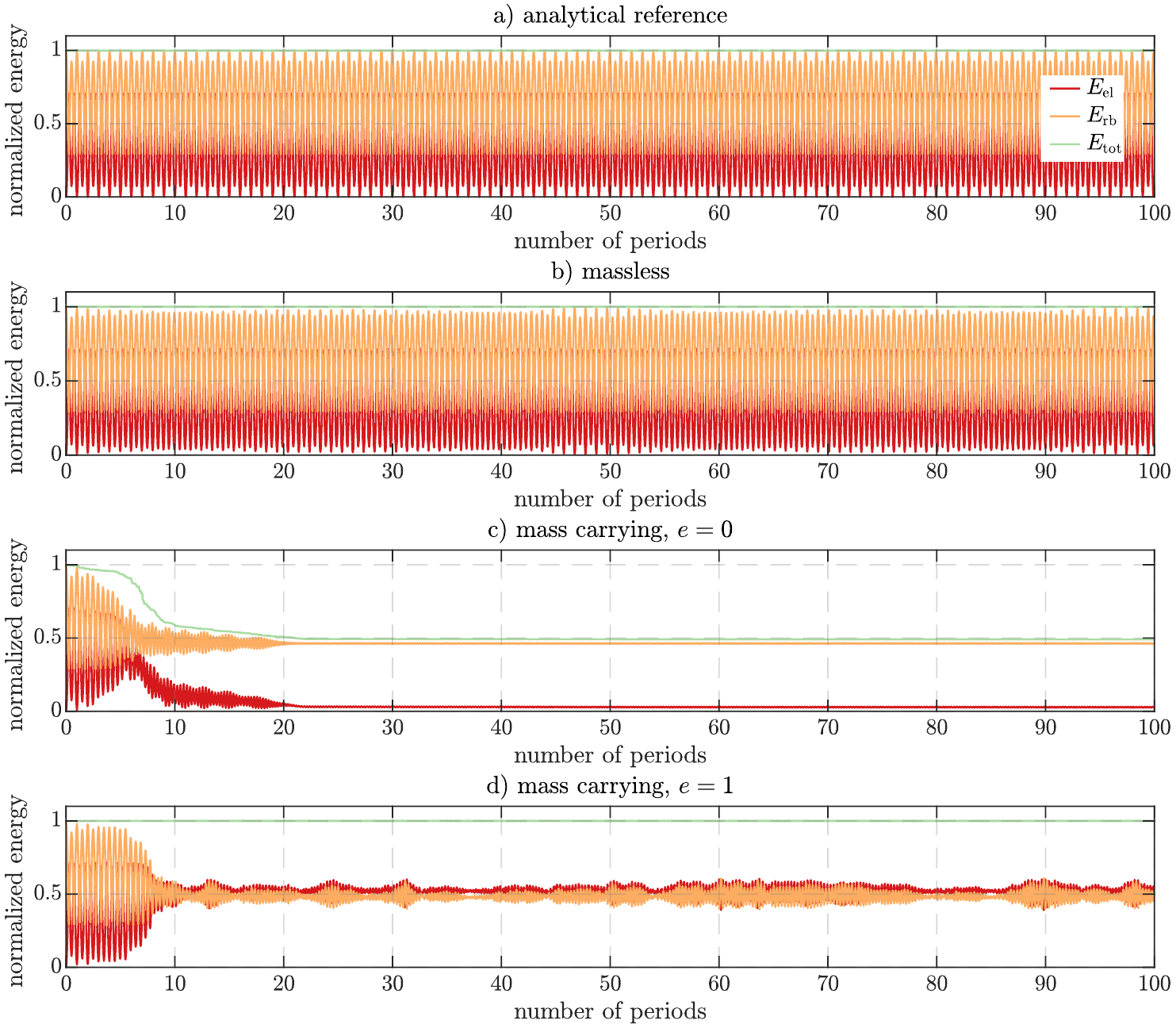}{Evolution of total energy $E_{\mathrm{tot}}$, energy in rigid body modes $E_{\mathrm{rb}}$ and energy in elastic modes (vibration energy) $E_{\mathrm{el}}$ over a longer time range; from top to bottom: exact solution, massless Craig-Bampton method, standard Craig-Bampton method with coefficient of restitution $\CoR=0$, standard Craig-Bampton method with coefficient of restitution $\CoR=1$}{1.0}
The time evolution of the boundary displacement $q_{\mathrm b}$ is depicted in \fref{ex1_plot_newTIdisp_dt4_DISP}.
Both numerical methods agree well with the exact reference during the first few bounces.
The discrepancy grows with time.
The proposed massless boundary approach predicts the long-term behavior of recurrent bounces with comparable height in a qualitatively correct way.
Apparently, a small deviation in terms of the time period of the bouncing process is present, which we attribute to the removed mass.
In contrast to the proposed approach, the mass-carrying boundary model does not predict the long-term bouncing behavior correctly; after some time, the boundary node does not lift up significantly anymore.
\\
\fref{ex1_plot_newTIdisp_dt4_ENER} depicts the time evolution of the total energy, $E_{\mathrm{tot}}$, and the individual contributions of energy contained in rigid body modes, $E_{\mathrm{rb}}$, and energy contained in elastic modes (vibration energy), $E_{\mathrm{el}}$.
Setting the restitution coefficient to $\CoR=0$ leads to artificial loss of energy.
At each impact, the kinetic energy of the boundary node is erased.
The energy decay goes on until there is no energy left for the dynamic process (\fref{ex1_plot_newTIdisp_dt4_ENER}c).
The remaining potential energy in the rigid body modes has to be attributed to the zero level of the potential energy in the gravity field, which is arbitrarily set to $x=0$.
Even in the static limit case, this energy would be larger than zero since the center of gravity is located at about $x=\ell/2$ (actually slightly below this, as the bar is compressed by its gravity load).
For $\CoR=1$, the total energy is conserved.
However, the finite mass at the boundary gives rise to spurious high-frequency oscillations, while the continuum model of the bar is in permanent contact for a finite time.
Consequently, the energy is rapidly transferred to elastic energy (vibrations).
\\
The results were also computed for $\Delta t = 10^{-3}$ and $\Delta t = 10^{-2}$, corresponding to a Courant number of $3$ and $30$, respectively.
The mass-carrying boundary approach became numerically unstable for the latter case.
The proposed massless boundary approach remained numerically stable and still showed the same qualitative behavior (recurrent bouncing with similar height).
However, the agreement with regard to the displacement time evolution was notably worse.

\subsection{Forced response of three-dimensional plate under frictionless contact} \label{sub:forced_response_of_a_cantilevered_3d_fe_beam}
Next, we consider a three-dimensional cantilevered plate under concentrated external loading, $f_{\mathrm{ex}}(t)$, subjected to frictionless unilateral contact with a rigid wall as illustrated in \fref{ex2_fig_beam3D01}.
The dimensions of the plate are $8~\mathrm{mm}$ x $40~\mathrm{mm}$ x $150~\mathrm{mm}$.
The material properties are $\rho = 8,220~\mathrm{kg}/\mathrm{m}^3$, $E=184~\mathrm{GPa}$ and $\nu = 0.33$.
The body is discretized with 8 x 20 x 101 C3D8 finite elements (8-node hexahedral elements with linear shape functions).
Contact is considered at three boundary nodes on one edge of the free end (\fref{ex2_fig_beam3D01}).
The asymmetry of the contact constraints ensures that torsion dynamics will be relevant besides bending dynamics.
The initial gap for all contact nodes is $g_0=0.1~\mathrm{mm}$.
In this example, we use the MacNeal method to derive the reduced-order model, retaining the $20$ lowest-frequency free-interface normal modes and the three residual flexibility attachment modes associated with the displacements in the contact normal direction ($y$-direction) of the three boundary nodes.
\fig[h!]{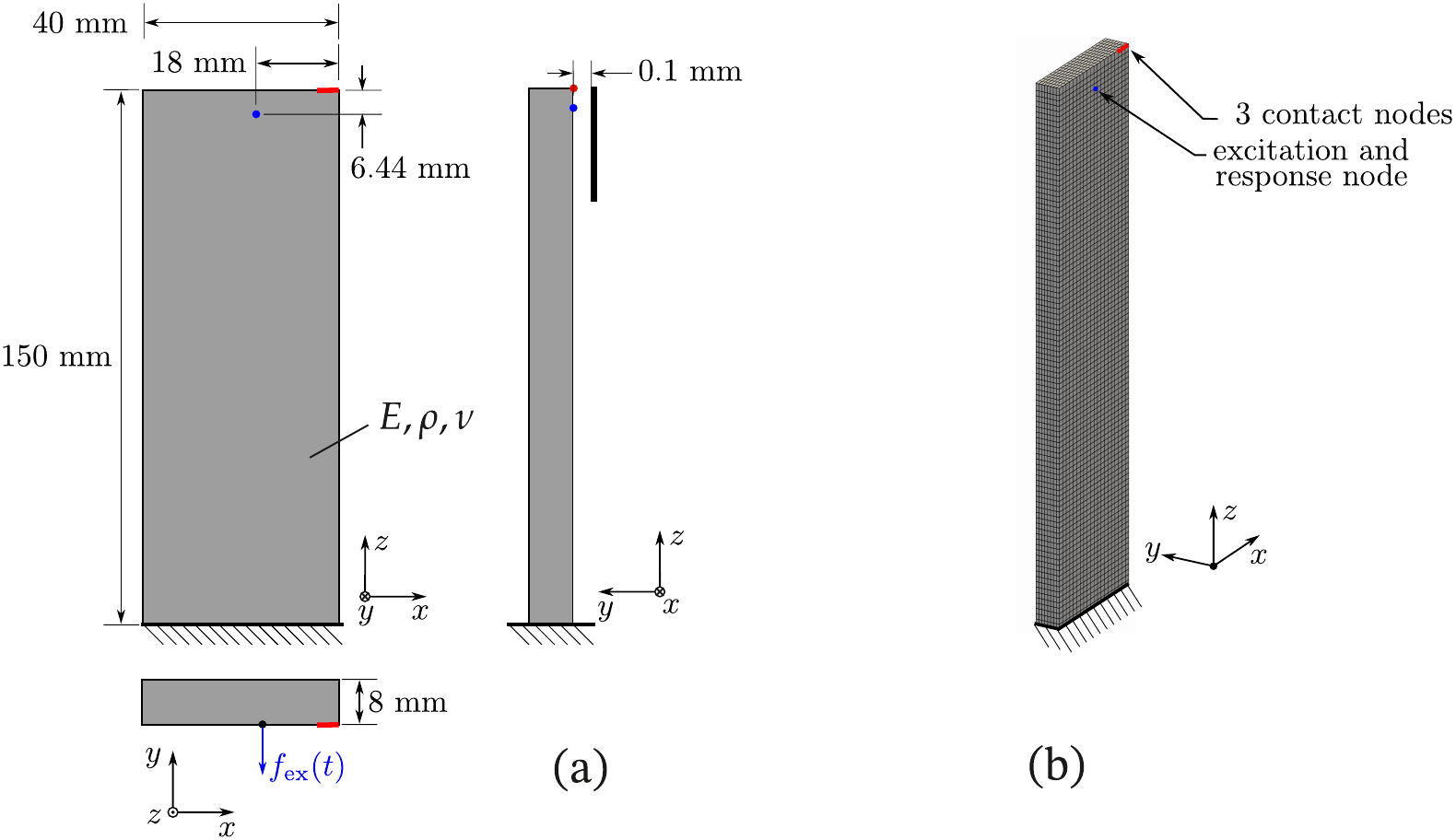}{Three-dimensional plate under frictionless contact: (a) schematic illustration of geometry and contact configuration, (b) finite element model}{1.0}
\\
We analyze the frequency response to a harmonic excitation with an amplitude of $1~\mathrm{N}$ and a frequency in the range near the lowest-frequency (flap-wise) bending mode.
A modal damping ratio of $1\%$ was specified for all free-interface modes.
We simulate the response to forward and backward frequency sweeps, and the response to constant-frequency excitation at some representative frequencies.
For the sweeps, the rate is such that the excitation frequency increases linearly with time by $1.5~\%$ in 100 pseudo-periods, where a pseudo-period is defined as the period corresponding to the considered natural frequency in the linear case of open contact.
In the time step integration, 500 time levels per pseudo-period were found to yield a good compromise between accuracy and efficiency.
For this benchmark, we compare the proposed time integrator with a nonlinear frequency-domain approach.
More specifically, we compute the periodic steady-state vibrations using the high-order harmonic balance method in conjunction with a Dynamic Lagrangian formulation to deal with the unilateral contact (without the need for regularization).
The method is described in \ref{asec:HBDL}.
It was implemented in the open source Matlab tool NLvib \cite{Krack.2019}.
For various values of the Dynamic Lagrangian parameter in the range $\varepsilon_{\mathrm{DL}}\in \left[10^4,10^7\right]$, the method converged to practically identical results.
The nominal harmonic truncation order was set to $H=20$, which was found to provide reasonable agreement with respect to the displacement and velocity of the response coordinate.
\\
An overview of the frequency response is given in \fref{ex2_plot_beam_FRF_sweep_all_relative} with respect to the negative $y$-coordinate of the response node, $q_{\mathrm R}$, indicated in \fref{ex2_fig_beam3D01}.
Away from resonance, the contact is open, the system behaves linearly.
For sufficiently large vibrations, contact interactions occur.
The temporarily closed contact provides positive stiffness, in time-average, giving rise to a hardening nonlinearity.
Moreover, the unilateral interactions limit the displacement in the direction of the wall.
In contrast, the displacements in the opposite direction increase further, leading to asymmetric vibrations with a non-zero mean value.
Due to the hardening nonlinearity and the light damping, the amplitude-frequency curve is bent towards the right, giving rise to two turning points connected by an overhanging branch.
As is well-known, the response corresponding to the overhanging branch is unstable.
The upper and lower branch of periodic responses are largely stable.
Looking very closely, one can find indications of a small window of non-periodic responses during the forward sweep.
As expected, jump phenomena are encountered during forward and backward sweeps near the respective turning point.
An excellent agreement between harmonic balance and time integration can be ascertained for $H=20$.
This agreement holds also for the phase projections depicted in \fref{HBvsTI_Harmonic_convergence} in the $u_{\mathrm R}$-$q_{\mathrm R}$ plane, which were obtained for fixed excitation frequency.
\fig[h!]{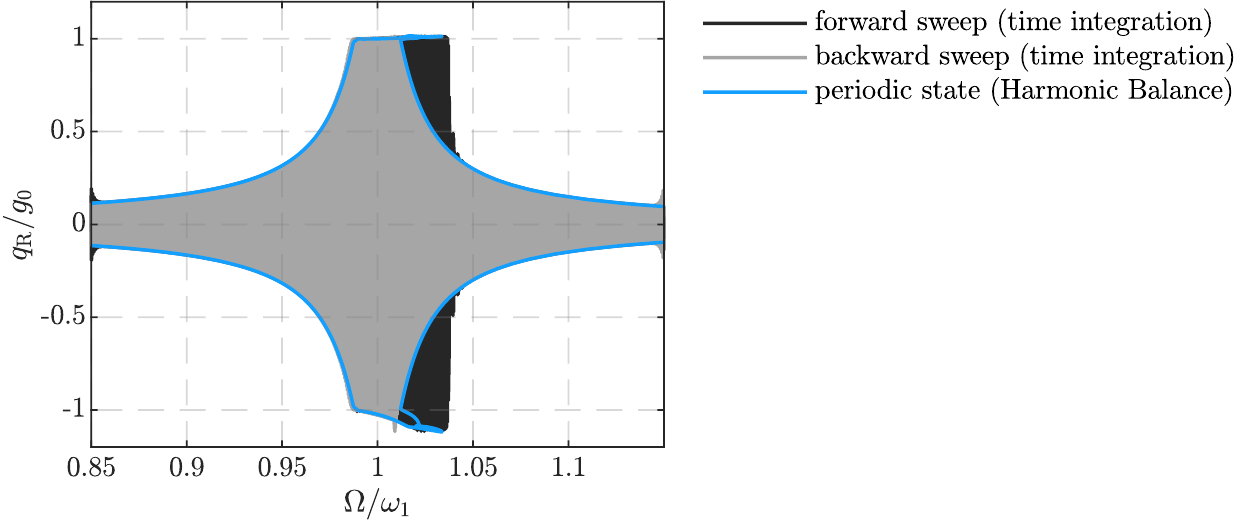}{Frequency response of the plate under frictionless contact}{1.0}
\fig[h!]{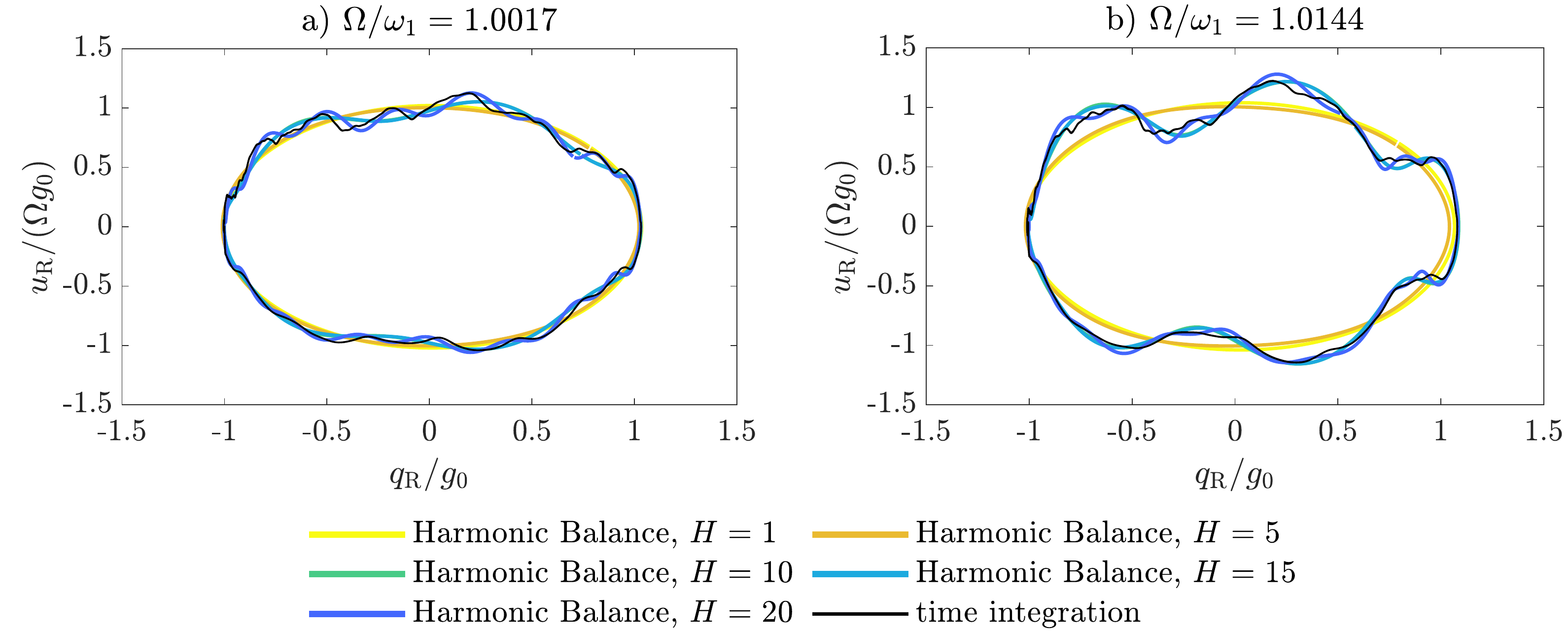}{Phase projections of periodic steady-state vibration response for time-constant excitation frequency}{1.0}
\\
As there is no rigorous theory available for the considered problem class of non-smooth, high-dimensional models, various method parameters had to be specified based on experience.
In particular, numerous different settings of the nominal step length in the pseudo arc-length continuation, the preconditioning properties, the solver tolerances and the number of time levels per period considered in the alternating frequency-time scheme were tested before the entire depicted solution branch was successfully computed.
A number of $N=2^{12}$ time levels (samples) per period was found to be necessary.
This is about 8 times larger than the number of time levels per period in the time step integration.
Apparently, this is due to non-smooth contact interactions, where the Fourier-based harmonic balance method suffers from the Gibbs phenomenon.
Consequently, a high number of significant harmonics is present in the nonlinear forces.
Thus, a relatively high number of samples is needed to reduce the inevitable aliasing errors, which may otherwise lead to convergence problems of the Newton-like solver.
In contrast to harmonic balance, the time integrator does not suffer from aliasing.
The avoidance of the spurious oscillations (Gibbs phenomenon) permits relatively large time steps.
The computational difficulties of harmonic balance for this benchmark naturally lead to a high computational effort.
Harmonic balance required about 90 minutes to compute the depicted frequency response branch.
Compared to this, the time integration of a sweep (either forward or backward) took only about 4 minutes.
Based on further investigations on the same benchmark, considering different excitation levels, different resonances, different method parameters, we are convinced that these orders of magnitude of the computational effort are representative.
This result is particularly interesting, because harmonic balance is usually orders of magnitude faster than time integration \cite{Krack.2019}.
For vibro-impact processes, apparently, this is not necessarily the case.

\subsection{Rotating blade in frictional contact with oval casing} \label{sec:blade_casing_rubbing}
Finally, we consider a rotating turbomachinery blade subjected to frictional impacts.
The technical motivation for this is that the clearance between blade tip and casing is designed to be as small as possible to improve the aerodynamic performance.
An abradable coating is commonly introduced.
However, if this is removed, the blade tip might come into contact with the underlying metallic casing (\emph{blade-casing rubbing}).
The accurate prediction of this behavior is computationally demanding because of the degree of detail needed both for modeling the bodies and the contact interactions \cite{legr2012a,Guerin.2018,Thorin.2018}.
We consider the benchmark problem illustrated in \fref{ex3_fig_bladecase_singleblade_02}.
The flexibility of the disk is regarded as negligible compared to that of the blades, and the coupling among blades is neglected, so that it is sufficient to model a single blade.
The geometry of the compressor blade corresponds to the NASA Rotor 37 and the finite element model is publicly available \cite{Piollet.2019}.
The blade is described with C3D15 finite elements (pyramids with mid-nodes, quadratic shape functions), leading to about 62,500 nodal degrees of freedom.
The material properties of the blade are $\rho=9,000~\mathrm{kg}/\mathrm{m}^3$, $E=210~\mathrm{GPa}$, and $\nu=0.3$.
The metallic casing is considered as rigid, which is a common simplification.
An oval geometry of the casing is considered, having a mean radius of about $317~\mathrm{mm}$, plus a sinusoidal perturbation with two waves around circumference.
The resulting tip clearance is specified as
\e{
	g_0(t)/\mathrm{mm} = 0.356 + 0.37 \cos(2\Omega_{\mathrm{rot}} t)\fp \label{eq:g0ovalcasing}
}
Consequently, two contact events per revolution are expected.
The angular velocity $\Omega_{\mathrm{rot}}$ is set to $\Omega_{\mathrm{rot}}=\frac{\omega_1}2$, with the lowest natural frequency $\omega_1/(2\pi)$ being $312.12~\mathrm{Hz}$.
Contact is considered at the 16 nodes indicated in \fref{ex3_fig_bladecase_singleblade_02} right. 
In the contact normal direction, unilateral interactions are modeled in terms of the Signorini law.
In the tangential contact plane, dry friction is modeled in terms of the spatial Coulomb law with a friction coefficient of $\mu=0.15$.
The rotation of the blade relative to the casing is considered by an imposed velocity in the respective tangential direction.
The Rubin method is commonly used in state-of-the-art simulations of blade-casing contact interactions \cite{legr2012a,bata2007}.
Hence, we use the Rubin method and the proposed MacNeal method to derive a mass-carrying and a massless boundary reduced order model, respectively.
Here, the 50 lowest-frequency free-interface normal modes are retained, along with the $3\cdot16=48$ residual flexibility attachment modes associated with all three degrees of freedom of the 16 contact nodes.
A modal damping ratio of $D=0.5\%$ is specified for all retained free-interface modes.
The simulation is carried out using the time step integration schemes described in \arefo{SymMoreau} and \sref{timestepping}, respectively.
In the mass-carrying case, the restitution coefficient is set to $\CoR=0.99$.
For $\CoR=1.0$, we did not find any method parameters that ensured numerical stability and provided reasonable accuracy.
It should be remarked that $\CoR=0$ is also a common choice for blade-casing interactions when using Moreau-like integrators \cite{Guerin.2018,Thorin.2018}.
We prefer a value closer to unity in order to avoid non-physical loss of energy.
\fig[h!]{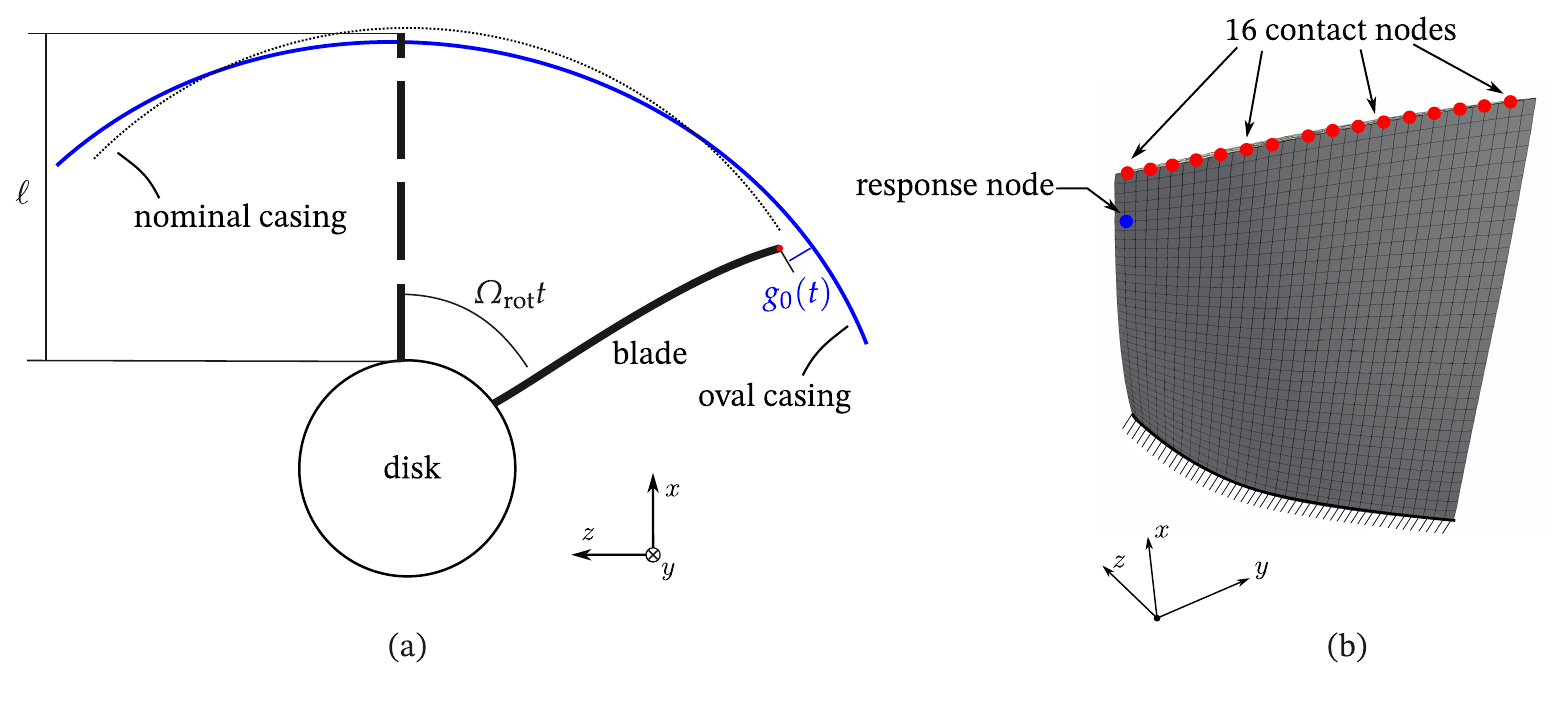}
{Rotating blade in frictional contact with oval casing: (a) schematic illustration of geometry and contact configuration, (b) finite element model with contact and response nodes}{1.0}
\\
The time evolution of the radial displacement at the response node, $q_{\mathrm{R}}$, is depicted in \fref{ex3_COMP_REFoverviewNtd50000_RELATIVE} for the case of zero initial displacement and velocity.
Here, a rather fine time discretization with $N=50,000$ time levels per revolution was used, corresponding to $\Delta t=2\pi/(N\Omega_{\mathrm{rot}})$.
Massless and mass-carrying boundary models are in excellent agreement.
\fig[h!]{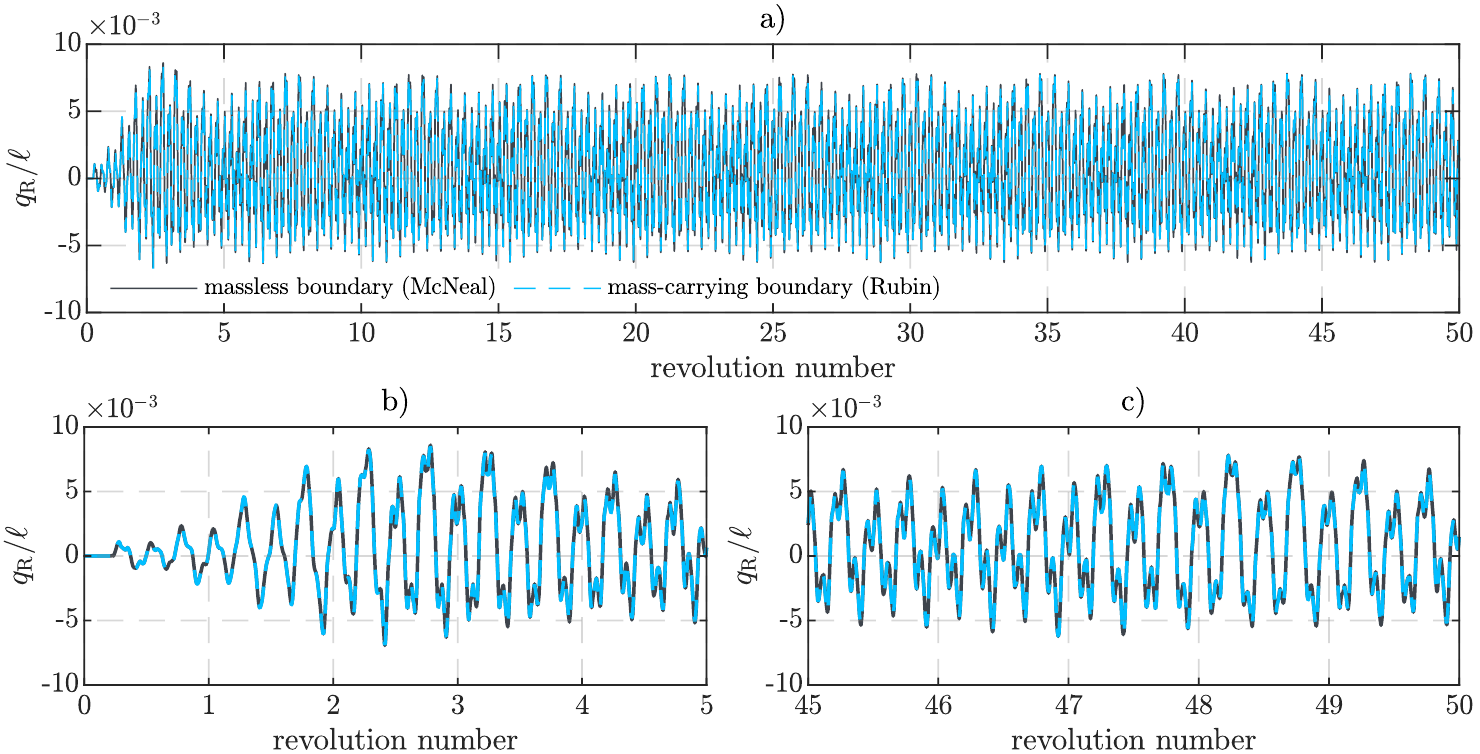}
{Time evolution of the response coordinate, : (a) overview, (b) zoom into initial phase, (c) zoom into steady-state phase; simulation with $N=50,000$ time levels per revolution}{1.0}
\\
Next, we analyze the convergence behavior of both numerical methods with the number of time levels.
As error measure, we use the relative root-mean-square deviation,
\e{
\varepsilon_{\mathrm{RMS}} = \sqrt{\frac{\sum\limits_{k} \left|q_{\mathrm R}^k -
q_{\mathrm R,\mathrm{ref}}^k\right|^2}{\sum\limits_{k} \left|q_{\mathrm R,\mathrm{ref}}^k\right|^2}}\fp
\label{eq:erms}
}
As reference, $q_{\mathrm R,\mathrm{ref}}$, the result for $N=50,000$ time levels per revolution is considered.
Here, the \emph{respective reference} is taken, \ie, the result of the massless boundary method is compared to the result of the massless boundary method with finest time discretization, and analogously for the mass-carrying boundary method.
We limit the analysis to the time span of the first five revolutions as illustrated in \fref{ex3_COMP_REFoverviewNtd50000_RELATIVE}b.
The results are depicted in \fref{ex3_COMP_performance}a.
\fig[h!]{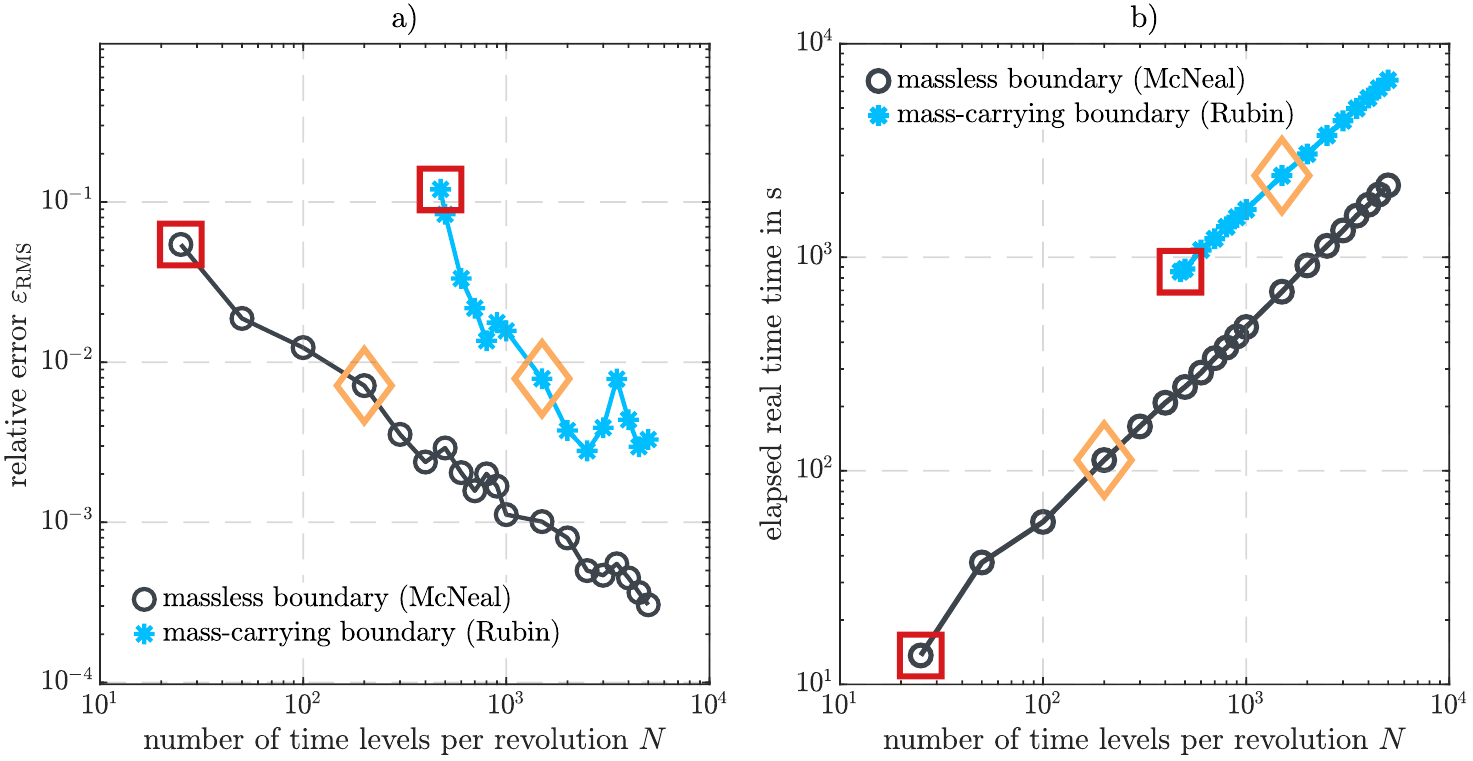}
{Comparison of the massless and mass-carrying boundary methods in terms of convergence and computational effort: (a) error vs. time levels per revolution, (b) computational effort vs. time levels per revolution.
	The red square and the orange diamond correspond to the low and the medium $N$ depicted in \fref{ex3_COMP_3lines_BOTH}, respectively.
}{1.0}
\fig[h!]{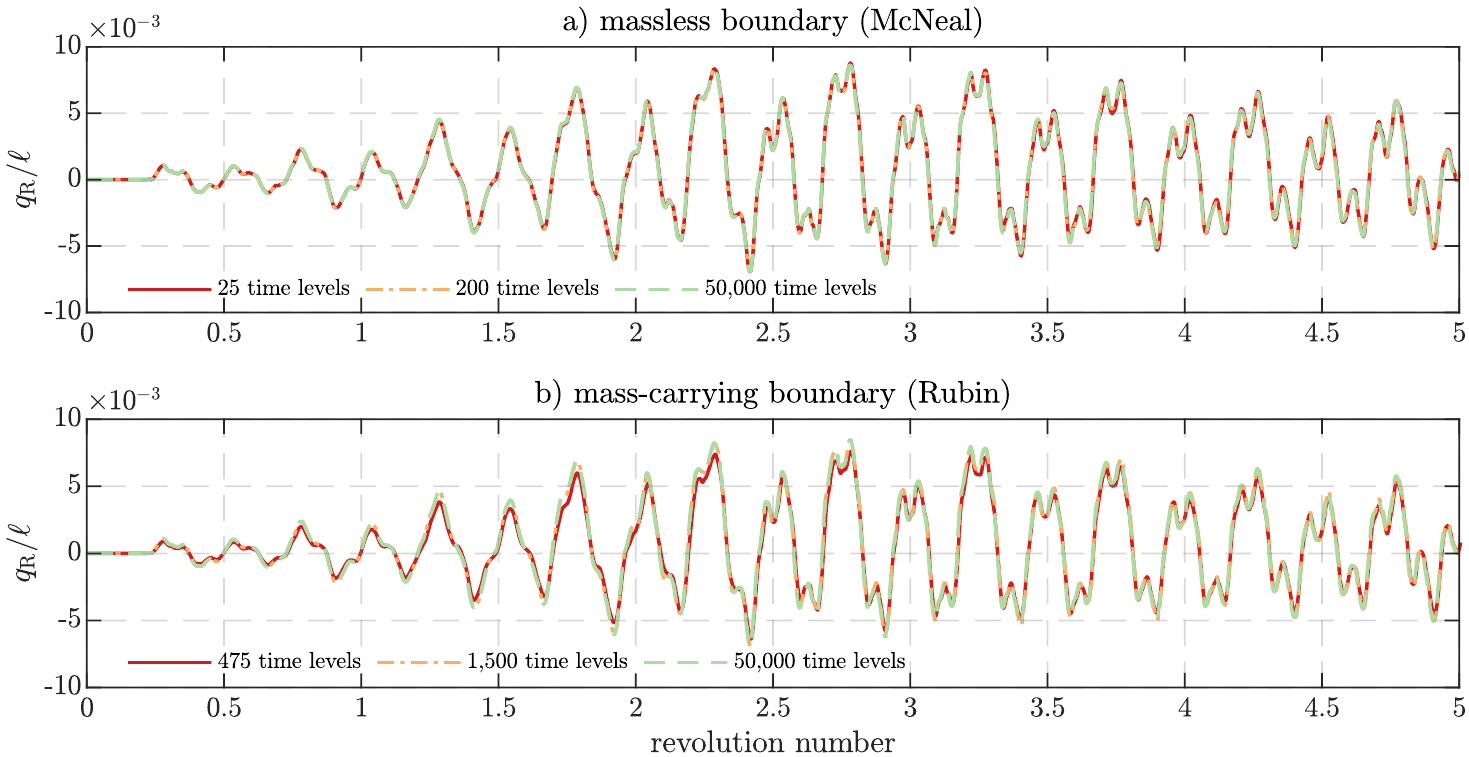}
{Time evolution of the response coordinate for the initial phase: (a) massless boundary method, (b) mass-carrying boundary method. Results are shown for the largest stable time step (red, solid), largest time step leading to an error $\varepsilon_{\mathrm{RMS}}<1~\%$ (orange dash-dotted), and the reference (green, dashed).
}{1.0}
%
The respective first point in the graphs (marked by a red square) corresponds to the largest (numerically) stable time step.
For the massless boundary method, $N=25$ is sufficient, while for the mass-carrying boundary method, $N=475$ time levels per revolution are needed.
While the error initially decays more rapidly in the mass-carrying boundary case, the error does not further decrease below $\varepsilon_{\mathrm{RMS}}\approx 8\cdot 10^{-3}$.
We attribute this to a slight numerical instability on the short time scale, associated with the spurious high-frequency oscillations.
In contrast, the overall error level is much lower in case of the massless boundary method and decays almost monotonously far below $\varepsilon_{\mathrm{RMS}}=10^{-3}$.
\\
The computational effort is depicted in \fref{ex3_COMP_performance}b in terms of wall time needed for the time span of the first 50 revolutions as illustrated in \fref{ex3_COMP_REFoverviewNtd50000_RELATIVE}a.
To ensure that the error is $\varepsilon_{\mathrm{RMS}}<1~\%$, the massless boundary method requires $N=200$, while the mass-carrying boundary method requires $N=1,500$ time levels per revolution.
The corresponding points are marked by orange diamonds in \fref{ex3_COMP_performance}.
The time evolution of the response coordinate is depicted in \fref{ex3_COMP_3lines_BOTH} for three different time steps: the largest stable time step, the time step leading to $\varepsilon_{\mathrm{RMS}}<1~\%$ and the finest time step (reference).
The results for $\varepsilon_{\mathrm{RMS}}<1~\%$ and the reference are indistinguishable for both methods.
The wall times for $\varepsilon_{\mathrm{RMS}}<1~\%$ are $110~\mathrm s$ in the case of the massless and $2400~\mathrm s$ in the case of the mass-carrying boundary method, corresponding to a speedup by a factor of more than 20.
As one may infer from \fref{ex3_COMP_3lines_BOTH}a, the result obtained using the massless boundary method for the largest stable time step ($N=25$) already gives excellent accuracy.
The speedup compared to the mass-carrying boundary method is then about 100.

\section{Conclusions}\label{sec:conclusions}
Overall, the proposed approach was shown to offer superior convergence and numerical robustness.
For the benchmarks considered in this work, the approach reduced the computational effort by about 1-2 orders of magnitude compared to the conventional mass-carrying boundary method.
In this comparison, the mass-carrying boundary method also relies on component mode synthesis and thus leads to the same mathematical model order.
We expect that the speedup over the initial finite element model is several orders of magnitude more, in spite of the loss of sparsity, but this still needs to be assessed.
We conclude that the approach is particularly well-suited for long-term simulations of elastodynamic contact problems, for instance vibro-impact processes.

\appendix
\setcounter{figure}{0}
\setcounter{table}{0}

\section{The Harmonic Balance method with a Dynamic Lagrangian approach to deal with unilateral contact} \label{asec:HBDL}
\nc{\naft}{N}
\nc{\rlinh}{\hat{\mm r}}
\nc{\td}[1]{\breve{#1}}
\nc{\gtd}{\td{\mm g}}
\nc{\qqbtd}{\td{\qq}_{\mathrm b}}
\nc{\epsAL}{\varepsilon_{\mathrm{AL}}}
\nc{\epsDL}{\varepsilon_{\mathrm{DL}}}
To make this article self-contained, we briefly recap here the Dynamic Lagrangian Harmonic Balance method \cite{naci2003}, used as reference in the numerical results section.
\\
Harmonic Balance seeks periodic solutions in the form of a Fourier series, truncated to order $H$,
\e{
	\qq \simeq \sum\limits_{h=-H}^{H} \hat{\qq}\left(h\right)~\ee^{\ii h\Omega t}\fp \label{eq:ansatz}
}
Herein, $\Omega$ is the fundamental angular oscillation frequency and $\hat{\square}\left(h\right)$ denotes the $h$-th complex Fourier coefficient of the quantity $\square$.
Since $\qq$ is real-valued, the Fourier coefficients form complex-conjugate pairs, $\hat{\qq}(-h) = \overline{\hat{\qq}(h)}$ for all $h\neq 0$, where $\overline{\square}$ denotes the complex conjugate.
This applies to all Fourier series considered in the following.
Harmonic Balance requires that the Fourier series of the residual, obtained by substituting \eref{ansatz} into \erefs{aeq}-\erefo{deq}, vanish up to order $H$.
This can be expressed as
\ea{
	\matrix{cc}{\mm K_{\mathrm{bb}} & \mm K_{\mathrm{bi}} \\ \mathrm{sym.} & \mm K_{\mathrm{ii}} - \left(h\Omega\right)^2\mm M_{\mathrm{ii}} + \ii h\Omega\mm D_{\mathrm{ii}}}~\vector{\hat{\qq}_{\mathrm b}\left(h\right) \\ \hat{\qq}_{\mathrm i}\left(h\right)} - \vector{\fexbh\left(h\right) \\ \fexih\left(h\right)} - \vector{\mm I \\ \mm 0}\hat{\mm\lambda}\left(h\right) = \mm 0 \quad h=-H,\ldots,H\fp
}
For simplicity, we replaced $\Wb$ by the identity matrix $\eye$ here and in the following.
As discussed in \sref{masslessCMS}, this is always possible by an appropriate coordinate transform under the given limitation to linear contact kinematics.
%
%
%
\\
The Fourier coefficients $\left\{\hat{\mm\lambda}(h)\right\}$ are determined by sampling the current estimate for $\qq$ at $\naft$ equidistant points along the period, then determining the corresponding $\mm\lambda$ in discrete time, and finally applying the discrete Fourier transform; \ie,
\ea{
	\hat{\mm g}\left(h\right) &= \hat{\mm q}_{\mathrm b}\left(h\right) + \mm g_0\left(h\right) \quad &h=-H,\ldots,H \label{eq:gh}\\
	\rlinh_{\mathrm b}\left(h\right) &= \mm K_{\mathrm{bb}} \hat{\mm q}_{\mathrm b}\left(h\right) + \mm K_{\mathrm{bi}}\hat{\mm q}_{\mathrm i}\left(h\right)-\fexbh\left(h\right) \quad &h=-H,\ldots,H \label{eq:rlin}\\
	\td{\mm r}_{\mathrm b}\left(k\right) &= \sum\limits_{h=-H}^{H} \rlinh_{\mathrm b}\left(h\right)~\ee^{\ii h\frac{2\pi}{\naft}k} \quad &k=0,\ldots,\naft-1 \label{eq:idft}\\
	\td{\mm\lambda}\left(k\right) &= \operatorname{proj}_{\mathbb R_0^+} \left(~\td{\mm r}_{\mathrm b}\left(k\right)~-~\epsDL\gtd\left(k\right)~\right) \quad &k=0,\ldots,\naft-1 \label{eq:projDL}\\
	\hat{\mm\lambda}\left(h\right) &= \frac{1}{\naft}\sum\limits_{k=0}^{\naft-1} \td{\mm\lambda}\left(k\right)~\ee^{-\ii h\frac{2\pi}{\naft}k} \quad &h=-H,\ldots,H \label{eq:dft}\fk
}
where $\left\{\gtd\left(k\right)\right\}$ is related to $\left\{\hat{\mm g}\left(h\right)\right\}$ via the inverse discrete Fourier transform as in \eref{idft}.


\section{Symmetric Moreau-like time integrator for conventional (mass-carrying boundary) models} \label{asec:SymMoreau}
To make this article self-contained, we briefly recap here the symmetric Moreau-like integrator \cite{Capobianco.2018}, used as reference in the numerical results section.
\\
In the conventional case of a mass-carrying boundary, no distinction is made between boundary and inner coordinates.
The time-discrete form of the measure differential equation and the contact laws can be expressed as
\ea{
	\mm M\left(\mm u^{k+\half}-\mm u^{k-\half}\right) + \mm D \frac{\mm u^{k+\half}+\mm u^{k-\half}}2 \Delta t + \mm K \mm q^{k} - \mm W~\Delta\mm P^k = \fex\left(t^k\right)\fk \label{eq:mditd}\\
	-\left(\mm\gamma^{k+\half}+\mm\CoR\mm\gamma^{k-\half}\right) \in \mathcal N_{\mathcal C}\left(\Delta\mm P^k\right)\fk \label{eq:mdiclaw}
}
where $\Delta\mm P$ are the contact percussions, defined as integral measures of both continuous contact forces and potential impulses within the considered discrete time interval.
$\mm\CoR$ is a diagonal matrix containing the corresponding normal and tangential coefficients of restitution.
Note that the admissible set $\mathcal C$ is the same as in the massless case as defined in \eref{admissibleset}.
It should be emphasized that the velocity $\mm u^k$ is approximated by the average value $\left(\mm u^{k+\half}+\mm u^{k-\half}\right)/2$ for the symmetric Moreau-like integrator.
The common (non-symmetric) case is recovered by replacing this with $\mm u^{k-\half}$.
\\
The algorithm is:
\begin{enumerate}
	\item Set $k=1$, $t^1=t_{\mathrm{start}}$ and initial $\qq^1$, $\uu^{\half}$.
	\item Solve \eref{mditd}, \eref{mdiclaw} and \eref{ckinematics} with respect to $\uu^{k+\half}$, $\Delta\mm P^k$.
	\item Evaluate $\mm q^{k+1} = \mm q^{k} + \uu^{k+\half}\Delta t$.
	\item If $t^k<t_{\mathrm{end}}$, increase $k$ by one and go back to step 2.
\end{enumerate}
To determine the active set of constraints, the contact kinematics in \eref{ckinematics} is evaluated at $t^k$.
Contacts having non-positive normal gap are considered active when computing $\uu^{k+\half}$, $\Delta\mm P^k$.
Step 2 leads to the implicit algebraic inclusion
\ea{
	-\left(\mm G\Delta\mm P^k + \mm c\right) ~\in~ \mathcal N_{\mathcal C}\left(\Delta\mm P^k\right) \label{eq:MMin}\\
	\text{with} \quad \mm G = \mm W\tra\left(\mm M + \frac{\Delta t}2\mm D\right)\inv\mm W\fk \\
	\mm c = \dot{\mm g}_0\left(t^k\right) + \mm\CoR\mm\gamma^{k-\half} + \mm W\tra\left(\mm M+\frac{\Delta t}2\mm D\right)\inv\left[\fex\left(t^k\right)-\mm K\qq^k-\left(\mm M-\frac{\Delta t} 2\mm D\right)\mm u^{k-\half}\right]\fp
}
Note that $\mm G$ inherits the symmetry and positive definiteness of $\mm D$ and $\mm M$.
In the case of a massless boundary, $\mm G$ is singular. 
Thus, the above described procedure is not applicable to massless boundary models. 
Instead, the algorithms developed in \sref{timestepping} should be used.

%

\section{Augmented Lagrangian approach to solve implicit algebraic inclusions} \label{asec:AL}
To make this article self-contained, we briefly describe here how we implemented the solution of the implicit algebraic inclusions in \fref{signalFlowPlan}, \eref{MMin}, and the projection in \eref{projDL}.
For the theoretical background and alternative techniques we refer to \cite{Acary.2008,Ascher.1998,Leine.2004,Studer.2009}.
\\
The implicit algebraic inclusions in \fref{signalFlowPlan} and \eref{MMin} take the form
\ea{
	-\left(\mm G\mm x + \mm c\right) \in \mathcal N_{\mathcal C} \left(\mm x\right)\fk \label{eq:iai}
}
with $\mm x = [\mm x_1;\ldots;\mm x_{\ncon}]$, $\mm x_j = [x_{\mathrm n,j};x_{\mathrm t1,j};x_{\mathrm t2,j}]$,
and $\mathcal C = \mathcal C_1 \times \ldots \times \mathcal C_{\ncon}$, $\mathcal C_j = \mathbb R_0^+\times \mathcal D\left(x_{\mathrm n,j}+x_{\mathrm n,j}^0\right)$.
\eref{iai} can be rewritten as (see \eg \cite{Acary.2008,Studer.2009})
\ea{
	\mm x = \operatorname{proj}_{\mathcal C}\left[~\mm x - \epsAL\left(\mm G\mm x+\mm c\right)~\right]\fp \label{eq:prox}
}
This can be interpreted as an Augmented Lagrangian approach \cite{Studer.2009}.
The larger $\epsAL$, the quicker the convergence typically.
However, a too large $\epsAL$ leads to divergence.
For the mass-carrying boundary methods, $\epsAL$ is set as proposed in \cite{Studer.2009}.
This provided optimal performance; in fact, doubling this value led to divergence for the benchmark in \sref{blade_casing_rubbing}.
The massless boundary method was found to be much less sensitive to the particular choice of $\epsAL$.
For the benchmark in \sref{blade_casing_rubbing}, a value of three times the one proposed in \cite{Studer.2009} was set.
\\
We use the projected Jacobi relaxation method to solve \eref{prox}, as this is known for its good convergence in the case of positive definite Delassus matrices $\mm G$ \cite{Studer.2009}.
Compared to the common projected Gauss-Seidel method, we found that the Jacobi method is more easy to vectorize in order to make use of fast multi-threaded linear algebra operations.
The projected relaxation algorithm involves projections on the subsets of $\mathcal C$, \ie, on $\mathbb R_0^+$ and $\mathcal D\left(r\right)$.
These projections can be explicitly written as
\ea{
	\operatorname{proj}_{\mathbb R_0^+}\left(\xi\right) = \begin{cases} \xi & \xi \geq 0\\ 0 & \xi <0\end{cases}\fk \label{eq:projR}\\
	\operatorname{proj}_{\mathcal D(r)}\left(\mm\xi\right) = \begin{cases} r\frac{\mm\xi}{\sqrt{\mm\xi\tra\mm\xi}} & \sqrt{\mm\xi\tra\mm\xi}>r\\ \mm \xi & \sqrt{\mm\xi\tra\mm\xi}\leq r\end{cases}\fp \label{eq:projD}
}
%


\begin{thebibliography}{10}

\bibitem{Babitsky.2013}
V.~I. Babitsky, {\em Theory of vibro-impact systems and applications}.
\newblock {Springer Science {\&} Business Media}, 2013.

\bibitem{Seifried.2010}
R.~Seifried, W.~Schiehlen, and P.~Eberhard, ``The role of the coefficient of
  restitution on impact problems in multi-body dynamics,'' {\em Proceedings of
  the IMechE}, vol.~224, no.~3, pp.~279--306, 2010.

\bibitem{wrig2006}
P.~Wriggers, {\em Computational Contact Mechanics}.
\newblock {Springer Berlin Heidelberg}, 2006.

\bibitem{john1989}
K.~L. Johnson, {\em Contact Mechanics}.
\newblock Cambridge: {Cambridge University Press}, 1989.

\bibitem{Ibrahim.2009}
R.~A. Ibrahim, {\em Vibro-Impact Dynamics}.
\newblock Berlin Heidelberg: Springer-Verlag, 2009.

\bibitem{klerk2008}
D.~de~Klerk, D.~J. Rixen, and S.~N. Voormeeren, ``General framework for dynamic
  substructuring: History, review and classification of techniques,'' {\em AIAA
  Journal}, vol.~46, no.~5, pp.~1169--1181, 2008.

\bibitem{carp1991}
N.~J. Carpenter, R.~L. Taylor, and M.~G. Katona, ``Lagrange constraints for
  transient finite element surface contact,'' {\em International Journal for
  Numerical Methods in Engineering}, vol.~32, no.~1, pp.~103--128, 1991.

\bibitem{Leine.2004}
R.~I. Leine and H.~Nijmeijer, {\em Dynamics and Bifurcations of Non-Smooth
  Mechanical Systems}, vol.~18 of {\em Lecture Notes in Applied and
  Computational Mechanics}.
\newblock {Springer Berlin}, 2004.

\bibitem{Acary.2008}
V.~Acary and B.~Brogliato, {\em Numerical methods for nonsmooth dynamical
  systems: applications in mechanics and electronics}.
\newblock {Springer Science {\&} Business Media}, 2008.

\bibitem{Acary.2013}
V.~Acary, ``Projected event-capturing time-stepping schemes for nonsmooth
  mechanical systems with unilateral contact and coulomb's friction,'' {\em
  Computer Methods in Applied Mechanics and Engineering}, vol.~256,
  pp.~224--250, 2013.

\bibitem{Tschigg.2018}
S.~Tschigg and R.~Seifried, ``Efficient impact analysis using reduced flexible
  multibody systems and contact submodels,'' in {\em 6th European Conference on
  Computational Mechanics: Solids, Structures and Coupled Problems, ECCM 2018
  and 7th European Conference on Computational Fluid Dynamics, ECFD 2018},
  pp.~2711--2722, 2018.

\bibitem{Krause.2009}
R.~Krause and M.~Walloth, ``A time discretization scheme based on rothe's
  method for dynamical contact problems with friction,'' {\em Computer Methods
  in Applied Mechanics and Engineering}, vol.~199, no.~1, pp.~1--19, 2009.

\bibitem{krause2012}
R.~Krause and M.~Walloth, ``Presentation and comparison of selected algorithms
  for dynamic contact based on the newmark scheme,'' {\em Selected Papers from
  NUMDIFF-12}, vol.~62, no.~10, pp.~1393--1410, 2012.

\bibitem{Gear.1985}
C.~W. Gear, B.~Leimkuhler, and G.~K. Gupta, ``Automatic integration of
  euler-lagrange equations with constraints,'' {\em Journal of Computational
  and Applied Mathematics}, vol.~12-13, pp.~77--90, 1985.

\bibitem{Chouly.2015}
F.~Chouly, P.~Hild, and Y.~Renard, ``A nitsche finite element method for
  dynamic contact: 2. stability of the schemes and numerical experiments,''
  {\em ESAIM: M2AN}, vol.~49, no.~2, pp.~503--528, 2015.

\bibitem{Ascher.1998}
{Uri M. Ascher} and {Linda R. Petzold}, ``Computer methods for ordinary
  differential equations and differential-algebraic equations,'' 1998.

\bibitem{Khenous.2008}
H.~B. Khenous, P.~Laborde, and Y.~Renard, ``Mass redistribution method for
  finite element contact problems in elastodynamics,'' {\em European Journal of
  Mechanics-A/Solids}, vol.~27, no.~5, pp.~918--932, 2008.

\bibitem{Hager.2008}
C.~Hager, S.~H{\"u}eber, and B.~I. Wohlmuth, ``A stable energy-conserving
  approach for frictional contact problems based on quadrature formulas,'' {\em
  International Journal for Numerical Methods in Engineering}, vol.~73, no.~2,
  pp.~205--225, 2008.

\bibitem{Hager.2009}
C.~Hager and B.~Wohlmuth, ``Analysis of a space-time discretization for dynamic
  elasticity problems based on mass-free surface elements,'' {\em SIAM J.
  Numer. Anal.}, vol.~47, no.~3, pp.~1863--1885, 2009.

\bibitem{Renard.2010}
Y.~Renard, ``The singular dynamic method for constrained second order
  hyperbolic equations: Application to dynamic contact problems,'' {\em Journal
  of Computational and Applied Mathematics}, vol.~234, no.~3, pp.~906--923,
  2010.

\bibitem{Tkachuk.2013}
A.~Tkachuk, B.~I. Wohlmuth, and M.~Bischoff, ``Hybrid-mixed discretization of
  elasto-dynamic contact problems using consistent singular mass matrices,''
  {\em International Journal for Numerical Methods in Engineering}, vol.~94,
  no.~5, 2013.

\bibitem{Hollkamp.2008}
J.~J. Hollkamp and R.~W. Gordon, ``Reduced-order models for nonlinear response
  prediction: Implicit condensation and expansion,'' {\em Journal of Sound and
  Vibration}, vol.~318, no.~4--5, pp.~1139--1153, 2008.

\bibitem{Kuether.2015b}
R.~J. Kuether, M.~S. Allen, and J.~J. Hollkamp, ``Modal substructuring of
  geometrically nonlinear finite-element models,'' {\em AIAA Journal}, vol.~54,
  no.~2, pp.~691--702, 2015.

\bibitem{Krack.2016}
M.~Krack, L.~Salles, and F.~Thouverez, ``Vibration prediction of bladed disks
  coupled by friction joints,'' {\em Archives of Computational Methods in
  Engineering}, vol.~24, no.~3, pp.~589--636, 2017.

\bibitem{Doyen.2011}
D.~Doyen, A.~Ern, and S.~Piperno, ``Time-integration schemes for the finite
  element dynamic signorini problem,'' {\em SIAM Journal on Scientific
  Computing}, vol.~33, no.~1, pp.~223--249, 2011.

\bibitem{Dabaghi.2014}
F.~Dabaghi, A.~Petrov, J.~Pousin, and Y.~Renard, ``Convergence of mass
  redistribution method for the one-dimensional wave equation with a unilateral
  constraint at the boundary,'' {\em ESAIM: M2AN}, vol.~48, no.~4,
  pp.~1147--1169, 2014.

\bibitem{Capobianco.2018}
G.~Capobianco and S.~R. Eugster, ``Time finite element based moreau-type
  integrators,'' {\em International Journal for Numerical Methods in
  Engineering}, vol.~114, no.~3, pp.~215--231, 2018.

\bibitem{rixe2004}
D.~J. Rixen, ``A dual craig--bampton method for dynamic substructuring,'' {\em
  Journal of Computational and Applied Mathematics}, vol.~168, no.~1-2,
  pp.~383--391, 2004.

\bibitem{macn1971}
R.~H. MacNeal, ``A hybrid method of component mode synthesis,'' {\em Computers
  {\&} Structures}, vol.~1, no.~4, pp.~581--601, 1971.

\bibitem{crai1968}
R.~R. Craig and M.~C. Bampton, ``Coupling of substructures for dynamic
  analysis,'' {\em AIAA Journal}, vol.~6, no.~7, pp.~1313--1319, 1968.

\bibitem{sher2013}
K.~Sherif, W.~Witteveen, H.~J. Holl, H.~Irschik, and K.~Mayrhofer, ``Effiziente
  simulation von~arbeitswalze und st{\"u}tzwalze mit ber{\"u}cksichtigung
  lokaler effekte.'' Paper-ID 219, Proc. of SIRM 2013 -- 10th International
  Conference on Vibrations in Rotating Machines, February 25-27, Berlin,
  Germany, 2013.

\bibitem{Sherif.2012}
K.~Sherif, W.~Witteveen, and K.~Mayrhofer, ``Quasi-static consideration of
  high-frequency modes for more efficient flexible multibody simulations,''
  {\em Acta Mechanica}, vol.~223, no.~6, pp.~1285--1305, 2012.

\bibitem{bata2007}
A.~Batailly, M.~Legrand, P.~Cartraud, C.~Pierre, and J.-P. Lombard, ``Study of
  component mode synthesis methods in a rotor-stator interaction case.''
  Proceedings of the ASME International Design Engineering Technical
  Conferences {\&} Computers and Information in Engineering Conference,
  September 4-7, Las Vegas, NV, USA, pp. 1--8, 2007.

\bibitem{Dabaghi.2016}
F.~Dabaghi, A.~Petrov, J.~Pousin, and Y.~Renard, ``A robust finite element
  redistribution approach for elastodynamic contact problems,'' {\em Applied
  Numerical Mathematics}, vol.~103, pp.~48--71, 2016.

\bibitem{Schreyer2016}
F.~Schreyer and R.~I. Leine, ``A mixed shooting -- harmonic balance method for
  unilaterally constrained mechanical systems,'' vol.~63, pp.~298--313, 2016.

\bibitem{Dabaghi.2019}
F.~Dabaghi, P.~Krej{\v{c}}{\'i}, A.~Petrov, J.~Pousin, and Y.~Renard, ``A
  weighted finite element mass redistribution method for dynamic contact
  problems,'' {\em Journal of Computational and Applied Mathematics}, vol.~345,
  pp.~338--356, 2019.

\bibitem{Krack.2019}
M.~Krack and J.~Gross, {\em Harmonic Balance for Nonlinear Vibration Problems}.
\newblock Springer, 2019.

\bibitem{legr2012a}
M.~Legrand, A.~Batailly, B.~Magnain, P.~Cartraud, and C.~Pierre, ``Full
  three-dimensional investigation of structural contact interactions in
  turbomachines,'' {\em Journal of Sound and Vibration}, vol.~331, no.~11,
  pp.~2578--2601, 2012.

\bibitem{Guerin.2018}
N.~Gu{\'e}rin, A.~Thorin, F.~Thouverez, M.~Legrand, and P.~Almeida,
  ``Thermomechanical model reduction for efficient simulations of rotor-stator
  contact interaction,'' {\em Journal of Engineering for Gas Turbines and
  Power}, vol.~141, no.~2, 2018.

\bibitem{Thorin.2018}
A.~Thorin, N.~Gu{\'e}rin, M.~Legrand, F.~Thouverez, and P.~Almeida, ``Nonsmooth
  thermoelastic simulations of blade--casing contact interactions,'' {\em
  Journal of Engineering for Gas Turbines and Power}, vol.~141, no.~2, 2018.

\bibitem{Piollet.2019}
E.~Piollet, F.~Nyssen, and A.~Batailly, ``Blade/casing rubbing interactions in
  aircraft engines: Numerical benchmark and design guidelines based on nasa
  rotor 37,'' {\em Journal of Sound and Vibration}, p.~114878, 2019.

\bibitem{naci2003}
S.~Nacivet, C.~Pierre, F.~Thouverez, and L.~Jezequel, ``A dynamic lagrangian
  frequency-time method for the vibration of dry-friction-damped systems,''
  {\em Journal of Sound and Vibration}, vol.~265, no.~1, pp.~201--219, 2003.

\bibitem{Studer.2009}
C.~Studer, {\em Numerics of unilateral contacts and friction: modeling and
  numerical time integration in non-smooth dynamics}.
\newblock {Springer Science {\&} Business Media}, 2009.

\end{thebibliography}

\end{document}